\documentclass{article}%

\usepackage{graphicx} %
\usepackage{hyperref}
\usepackage{amsmath}%
\usepackage{amssymb}%
\usepackage{amsthm}%
\usepackage{thmtools}
\usepackage{mathtools}%
\usepackage[shortlabels]{enumitem}
\usepackage[nameinlink]{cleveref}%
\usepackage[a4paper, margin=0.75in]{geometry}%
\usepackage{multirow}
\usepackage[numbers,sort&compress]{natbib} %

\declaretheorem[numberwithin=section]{theorem}
\declaretheorem[sibling=theorem]{
    corollary,
    conjecture,
    proposition,
    lemma,
    definition,
    example
}

\DeclarePairedDelimiter\ket{\lvert}{\rangle}
\DeclarePairedDelimiterX\braket[2]{\langle}{\rangle}{#1 \delimsize\vert #2}
\DeclarePairedDelimiterX\ketbra[2]{\delimsize\vert}{\delimsize\vert}{#1 \rangle \langle #2}

\newcommand{\Z}{\mathbb{Z}}

\newcommand{\R}{\mathbb{R}}

\newcommand{\Q}{\mathbb{Q}}
\newcommand{\C}{\mathbb{C}}

\newcommand{\inv}{^{-1}}

\newcommand{\kz}{\ket{0}}
\newcommand{\ko}{\ket{1}}

\newcommand{\kp}{\ket{+}}
\newcommand{\km}{\ket{-}}
\newcommand{\kpp}{\ket{++}}

\newcommand{\kpm}{\ket{\pm}}
\newcommand{\kmp}{\ket{\mp}}

\newcommand{\kpsi}{\ket{\psi}}

\newcommand{\onesqrttwo}{\frac{1}{\sqrt{2}}}
\newcommand{\onesqrtn}{\frac{1}{\sqrt{n}}}
\newcommand{\onesqrtm}{\frac{1}{\sqrt{m}}}

\newcommand{\allones}{\mathbf{1}}
\newcommand{\allonesn}{\mathbf{1}_n}
\newcommand{\allonesm}{\mathbf{1}_m}
\usepackage{xspace}

\newcommand{\ER}{Erd\H{o}s-R\'enyi\xspace}%

\DeclareMathOperator{\sign}{\operatorname{sign}}%

\title{Programmable Quantum-Like bits from Signed Regular Graphs}
\date{\today}

\usepackage{authblk}%
\author[1]{Ethan Dickey%
  \thanks{Corresponding author: 
  \href{mailto:dickeye@purdue.edu}{\texttt{dickeye@purdue.edu}}}}
\author[1]{Abhijeet Vyas}
\author[2]{Sabre Kais}
\affil[1]{Department of Computer Science, Purdue University, West Lafayette, IN 47907, USA}
\affil[2]{Department of Electrical and Computer Engineering and Department of Chemistry, North Carolina State University, Raleigh, NC 27606}
\providecommand{\keywords}[1]{\small \textbf{\textit{Keywords---}} #1}

\begin{document}
\maketitle

\begin{abstract}
    Extending upon observations of the emergence of quantum-like (QL) states from classical complex synchronized networks, this work adds mathematical rigor to the analysis of single QL bits constructed from adjacency-matrix eigenvectors.  First, we rigorously show that symmetric construction of such networks (regular undirected/symmetric bipartite graph $G_C$ connecting two regular undirected subgraphs $G_A,\,G_B$) leads to an equal superposition of the $\kp, \km$ Hadamard states (with basis $\kz,\,\ko$ set from eigenvectors of the subgraphs), and provide an analysis of sufficient conditions on the network for construction of such states.
    Second, we prove two methods to construct arbitrary single qubit states $\kpsi = a\kz + b\ko,\, |a|^2+|b|^2=1$, and give switching lemmas for their boundaries: (i) by detuning the two subgraphs regularities and (ii) by asymmetrically allowing the bipartite connection matrix $C$ to be directed and detuning those regularities. Although motivated by using complex synchronized networks for quantum information storage and computation, the proofs for these methods rely only on the structure of the graph embedded in the adjacency matrix. Thus, synchronization is unnecessary; QL bits arise when edge weights are unit (or near-unit) and subgraphs are regular.  Results on combinations of random k-regular graphs (more precisely \ER graphs) may be independently interesting.

\end{abstract}


\section{Introduction}\label{sec:intro}

Synchronized networks occur in a wide variety of classical systems ranging from schools of fish, interacting bacterial networks, and blinking fireflies to power grid fluctuations and mechanisms underpinning memory processes in the brain \cite{strogatz2000kuramoto,arenas2008synchronization,shahal2020violins,francheto2024synchronization}. A particularly striking everyday example of a complex synchronized network is pedestrians crossing a footbridge.  When the bridge is full or swaying, the complex network of pedestrians nonlinearly synchronizes into lockstep, leading to nontrivial oscillations in the bridge \cite{boccaletti2008synchronizeddynamics}.

Synchronized networks -- particularly those with dense or nearly regular connectivity -- remain stable under surprisingly large perturbations (e.g., random removal of up to half of the edges, provided the network remains connected and the coupling strength exceeds the master stability condition
\footnote{In essence, this means the coupling is strong enough, given the network's structure, that any small deviations between nodes die out, keeping the system synchronized. See also the overview in \cite{acharyya2025masterstabilityfunctionscomplex}.}
\cite{pecora1998master,barahona2002synchronization}).  Furthermore, their adjacency (or Laplacian) spectrum regularly exhibit a large spectral gap: one or more eigenvalues distinguishably separated from the rest of the spectrum.  With such a robust spectral gap, one can ponder if stable computation can be done with such emergent eigenvalues and associated eigenvectors.  Indeed, if such stability gives rise to emergent behaviors when these networks interact, it can allow insight into mechanisms underpinning such processes.  A particularly relevant example of this is helping understand mechanisms in the human brain, and whether those complex synchronized networks are performing some kind of computation viewable from emergent eigenvectors of their structure.

Recently, Scholes has discovered that such networks can give rise to \textit{quantum-like} behavior, in particular that interacting multiple carefully constructed complex synchronized networks gives rise to emergent coherent quantum states \cite{scholes2023largecoherentstates,scholes2024quantum}. By coupling multiple synchronized subnetworks in a medium-connectivity regime, the eigenvectors associated with emergent eigenvalues of the subgraphs interact in a way analogous to quantum coherence and interference.

Underpinning this behavior is the approximate regularity of the underlying graphs. For instance, in a random $k$-regular graph one has a unique top eigenvalue $\lambda_1=k$, while the second eigenvalue satisfies $\lambda_2 \le 2\sqrt{k-1} + o(1)$ with high probability by the Alon–Boppana and Friedman theorems \cite{alon1986eigenvalues,friedman2003relative,nilli1991onthe} (specifically that a random regular graph is ``almost Ramanujan'' with high probability), yielding a nonzero spectral gap $k-\lambda_2$ that protects the coherent mode from the spectral bulk.

Beyond idealized regular graphs, many natural and engineered networks are better modeled as \ER random graphs $G(n,p)$, with $n$ labeled vertices and each of the $\binom{n}{2}$ possible edges present independently with probability $p$ \cite{erdos1959random,bollobas1998randomgraphs}, otherwise known as a Bernoulli random graph  \cite{guionnet2021bernoulli,csardi2025igraph}.  In this model the degree of each node concentrates sharply around its mean $np$ (by Chernoff bounds), and for
    \[ p \ge \frac{\log n + \omega(1)}{n}, \]  %
the graph is connected with high probability.  Spectrally, the adjacency matrix of $G(n,p)$ has a top eigenvalue
    \[ \lambda_1 = np + o(n), \]
almost surely (with probability tending to 1 as $n$ tends to $\infty$), as originally proven by Juhasz (1978) \cite{juhasz1978spectrum}
\footnote{Proposition 2 gives $\lim\limits_{n\to \infty} \frac{\lambda_1}{n} = p \implies \lambda_1 = pn+o(n)$ for square symmetric matrices with $a_{ii}=0$ and $a_{ij} =$ Bernoulli random variables with probability $p$.  For a $k$-regular graph ($k<n$) with random edge deletions of probability $1-p$ (giving average degree $d = kp$; this can be equivalently understood as ``masking'' the Bernoulli random graph with a $k$-regular $\allones$s matrix), following the proof of Proposition 2, we modify the proof to say that for a given row $i$, the probability $\Pr\left(|\frac1k \protect\sum_{j=1}^n a_{ij}-p| > \delta \right)$ is exponentially small, yielding $\lim\limits_{k\to \infty} \frac{\lambda_1}{k} = p \text{ w.h.p.} \implies \lambda_1 = kp+o(n)$.
    This corrects a bug in the adaptation (of techniques in \cite{guionnet2021bernoulli}) of the Bernoulli random matrix decomposition in \cite{scholes2025graphsexcitons} to directly provide that $\lambda_1^B = np$ is the largest eigenvalue for Bernoulli random graphs and that $k$-regular graphs with random edge deletions have $\lambda_1 = kp$ w.h.p.  This holds in expectation for ensembles of Bernoulli random graphs trivially.}.
This follows the work in \cite{scholes2025graphsexcitons}, while the bulk of the spectrum lies in the interval
    \[ \lambda_{i\neq 1} \in \left[-2 \sqrt{np(1-p)} \,,\, 2 \sqrt{np(1-p)} \right] + o(\sqrt{n}) \]
almost surely by the semicircle law for dense graphs \cite{furedi1981eigenvalues}
\footnote{In \cite{furedi1981eigenvalues}, using Theorem 1, eqn. (6), take $\sigma = \sqrt{p(1-p)}$ for Bernoulli random variables $a_{ij}$, and $O(n^{1/3}\log n) \in o(\sqrt n)$ as $n \to \infty$, see also \cite{vu2005spectral} for improved bounds.},
so that a gap
    \begin{equation} \lambda_1 - \lambda_2 \ge np - 2\sqrt{np(1-p)} \label{eqn:spec_gap_ER_graphs} \end{equation}
opens up whenever $np \gg \sqrt{np}$ (as $o(n)-o(\sqrt{n})>0$, if we abuse notation for intuition).  Thus, even though \ER graphs are not exactly regular, they inherit a large spectral gap in the dense regime \cite{chung2003spectra}.
See \Cref{app:spect_gap} for further analysis of this particular spectral gap.

Refer to \cite{scholes2025graphsexcitons} for more background on the analysis of emergent eigenvalues of \ER graphs (Bernoulli random matrices).  For further discussion of the bounds of the spectrum, especially when $np \gg n^{2/3}$, see \cite{erdHos2013spectral,huang2024optimal,alt2021extremal}.  Also relevant are Weyl's Inequalities (or Weyl's interlacing properties) \cite{bhatia1997matrix,godsil2013algebraic}.

In summary, we can model many ``real-life'' networks as \ER (Bernoulli) random graphs and get similar results for the value of the top eigenvalue $\lambda_1$ and the robustness of the spectral gap $\lambda_1-\lambda_2$ as with random $k$-regular graphs.  Moving forward, this paper primarily considers random $k$-regular graphs with the goal of using the top eigenvalue's eigenvector to perform robust (due to the spectral gap) quantum-like operations using graph manipulations.

Returning to synchronized networks, this spectral separation underpins both robustness and coherence.  In the master‐stability framework \cite{pecora1998master,barahona2002synchronization}, the stability of the synchronous manifold depends on the ratio of the largest to second‐smallest eigenvalues of the coupling Laplacian.  A large gap guarantees that only the principal mode (associated with uniform synchronization) persists, while all orthogonal perturbations decay.  It is precisely this mechanism that both protects against massive connection loss in dense random topologies and allows coupled synchronized subnetworks to exhibit emergent, quantum‐like coherent modes when arranged as in Scholes et al.’s construction \cite{scholes2023largecoherentstates,scholes2024quantum}.  See also \cite{tang2014synchronization,gambuzza2021stability,strogatz2000kuramoto,arenas2008synchronization}.

More recently, similar studies showed that the product space of these regular graphs that model complex synchronized networks can robustly model a Hilbert space in a 1-1 mapping from the product basis of quantum states to the eigenstates of these graphs \cite{scholes2025productstates}. They go on to show how to encode quantum information in the steady state of a classical network \cite{amati2025encoding}, how to perform maps on the product space of these graphs that correspond to quantum gates and measurement \cite{amati2025qlbits}, and that, under certain conditions, the state space of classical synchronized networks evolve according to unitary dynamics \cite{scholes2025dynamics}.

Separating slightly from the physical motivation derived from complex synchronized networks, this work studies the graphs required to give rise to such Quantum-Like (QL) behavior from a computer science and mathematical perspective.  While the work in \cite{scholes2025productstates} provides a robust 1-1 mapping to the product basis of quantum states, explicit construction and manipulation of such states requires either taking the Cartesian product of two graphs or performing a unitary transformation on the state space of a QL-bit, which are both highly unintuitive and expensive operations, with the Cartesian product resulting in an exponentially larger graph.  We demonstrate how to explicitly construct an arbitrary QL-bit for any $\kpsi = a\kz + b\ko$, for $|a|^2+|b|^2 = 1$,
by manipulating well-understood graph properties and prove rigorously the existence of such states.

\section{Definitions}\label{sec:definitions}

\begin{figure}[ht!]%
    \centering
    \includegraphics[width=1.05\textwidth]{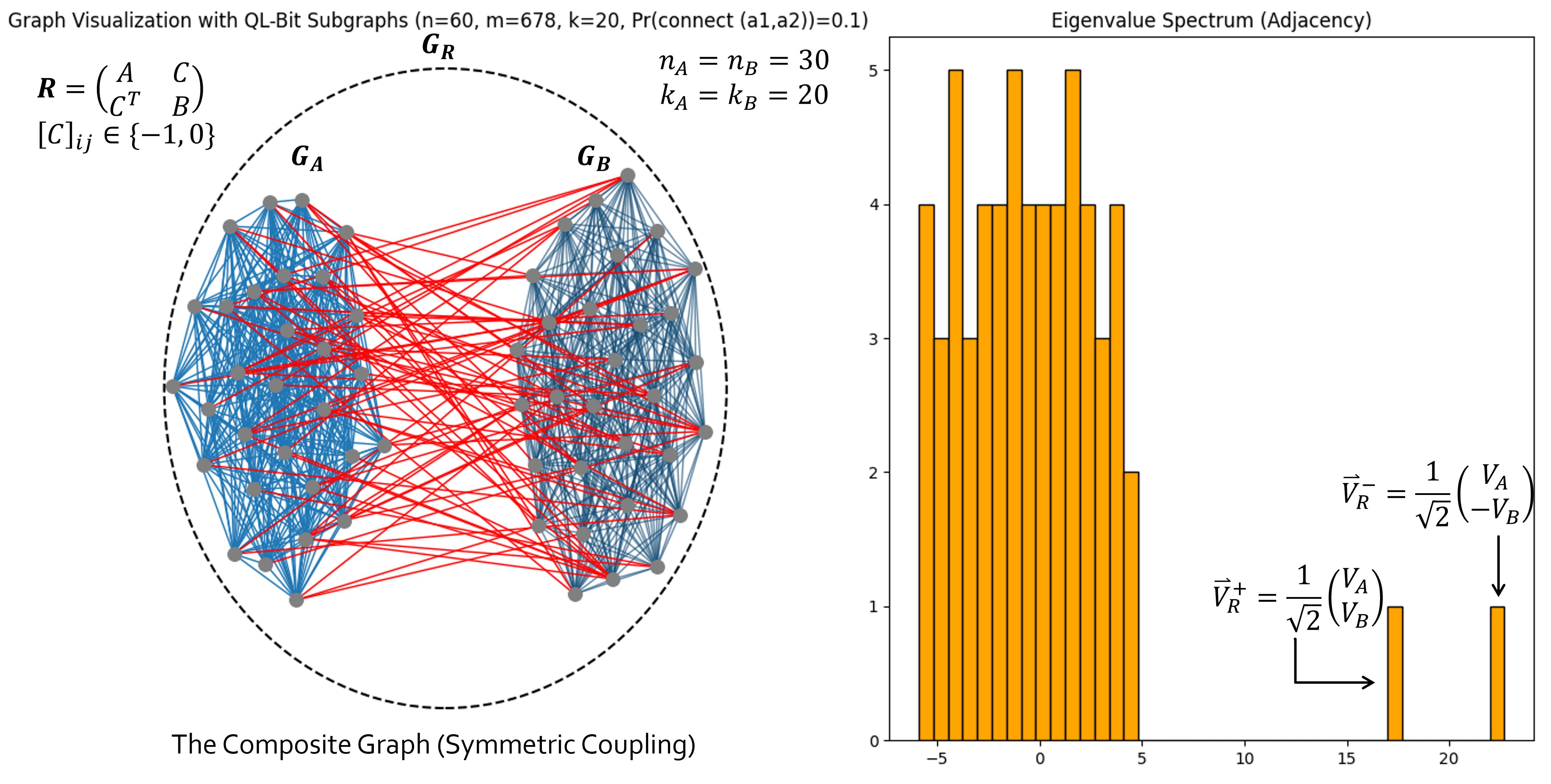}

    \caption{Regular graph and adjacency spectrum for the symmetric construction. Each subgraph has 30 nodes and regularity $k=20$. Each possible inter-subgraph edge of weight $-1$ is included independently with probability $0.1$, giving expected connecting regularity $l=3$. The realization shown has 678 total edges, and its two emergent eigenvalues lie near $k-l=17$ and $k+l=23$, consistent with \Cref{lem:basic_comp_eigvecs}.}
    \label{fig:graphandspectrum_adj_msubgraphNone}
\end{figure}

Let $G$ be a graph represented by its adjacency matrix. In our construction, a single QL--bit is built from two subgraphs $G_A$ and $G_B$ with adjacency matrices $A$ and $B$, respectively, as in \Cref{fig:graphandspectrum_adj_msubgraphNone}. Assume that:
\begin{itemize}
    \item $G_A$ is $k_A$--regular on $n$ vertices, and $G_B$ is $k_B$--regular on $m$ vertices.
    \item Their Perron--Frobenius eigenvectors \cite{pillai2005perron} are given by
    \begin{equation}\label{eqn:vavb_eigvecs}
        V_A = \onesqrtn\,\allonesn, \quad V_B = \onesqrtm\,\allonesm ,
    \end{equation}
    with associated eigenvalues
    \begin{equation}\label{eqn:vavb_eigvals}
        \lambda_A = k_A, \quad \lambda_B = k_B .
    \end{equation}
\end{itemize}

\noindent In the simplest (symmetric) construction, the two subgraphs are coupled by a bipartite connection matrix $C$ (of size $n\times m$, for $n$ rows) with entries in $\{-1,0\}$
    (see footnote\footnote{Due to motivation from complex synchronized networks, connecting edges are usually negative, $[C]_{ij} \in \{-1, 0\}$.  In the theoretical analysis, having negative connection edges $[C]_{ij} \in \{-1, 0\}$ only switches the top two (emergent) eigenvalues, putting the eigenvalue corresponding to $\km$ above the one corresponding to $\kp$. This is why the basis is chosen as $\kmp$ as opposed to the standard Hadamard basis $\kpm$.\label{footnote:neg_C}}),
so that the composite adjacency matrix is%
\footnote{$R$ is precisely the adjacency matrix of a signed graph, where each nonzero entry carries a $\pm 1$ sign. For a comprehensive survey of the adjacency spectra of signed graphs (including basic eigenvalue bounds, interlacing, and connections to two‐graphs), see \cite{belardo2018SignedGraphs}.}
    \[ R = \begin{pmatrix} A & C \\[1mm] C^T & B \end{pmatrix} . \]
When $C$ is chosen to be $l$--regular (i.e. every vertex in the appropriate partition has exactly $l$ connections with weight $-1$), one may define the \emph{quantum--like} basis vectors as
\begin{equation}\label{eqn:kp_km}%
    \psi_+ \equiv \kp = \onesqrttwo\left(\begin{pmatrix} V_A \\[1mm] \mathbf{0}_m \end{pmatrix} + \begin{pmatrix} \mathbf{0}_n \\[1mm] V_B \end{pmatrix} \right) = \onesqrttwo\begin{pmatrix} V_A \\[1mm] V_B \end{pmatrix}, \quad
    \psi_- \equiv \km = \onesqrttwo\begin{pmatrix} V_A \\[1mm] -V_B \end{pmatrix} .
\end{equation}
We note here that the choice of the $\kpm$ basis (the Hadamard basis) allows for simplified analysis of the emergent eigenvectors. Equivalently, one may use the computational basis
    \[
        \kz \coloneqq \begin{pmatrix} V_A \\[1mm] \mathbf{0}_m \end{pmatrix}, \qquad
        \ko \coloneqq \begin{pmatrix} \mathbf{0}_n \\[1mm] V_B \end{pmatrix},
    \]
for which $\kp=\onesqrttwo(\kz+\ko)$ and $\km=\onesqrttwo(\kz-\ko)$.

The goal is to construct an emergent eigenvector of $R$ corresponding to an arbitrary state
    \[ \psi = a\,\psi_+ + b\,\psi_-,\quad\text{with}\quad |a|^2+|b|^2=1 . \]
In our treatment, we shall consider two methods for tuning the state:
\begin{enumerate}
    \item Varying the regularities $k_A$ and $k_B$ (``detuning'') in the symmetric coupling, and
    \item Allowing the off--diagonal coupling ($C$) to be directed (asymmetric) so that the two coupling blocks differ.
\end{enumerate}
We also consider allowing edge weights continuously in $[-1, 1]$ rather than in $\{\pm 1, 0\}$ in \Cref{sec:real_valued_C}, although this turns out to be not well physically motivated and trivially all-powerful.

\section{Symmetric Construction and Detuning}\label{sec:sym}

The value of the top eigenvalues of $R$ were previously only noted as observations in \cite{amati2025qlbits}.  Here, we rigorously prove their existence and thereby show their exact construction.

\begin{lemma}[Composition of Eigenvectors of $R$]\label{lem:basic_comp_eigvecs}
    Let $G_R$ be constructed from two disjoint $k$-regular subgraphs $G_A$ and $G_B$, each of order $n$, with entries $[A]_{ij}\,, [B]_{ij} \in\{0, 1\}$, coupled by an $l$-regular bipartite graph $G_C$ with undirected connection matrix $C$ (with $[C]_{ij}\in\{-1,0\}$). If $V_A$ and $V_B$ are the normalized eigenvectors corresponding to the largest eigenvalues of $A$ and $B$, respectively, as in \Cref{eqn:vavb_eigvecs}, then the vectors
        \[ V_R^- = \onesqrttwo\begin{pmatrix} V_A \\[1mm] -V_B \end{pmatrix} = \km \quad \text{and} \quad
           V_R^+ = \onesqrttwo\begin{pmatrix} V_A \\[1mm]  V_B \end{pmatrix} = \kp \]
    are eigenvectors of 
        \[ R=\begin{pmatrix} A & C \\[1mm] C^T & B \end{pmatrix} \]
    with eigenvalues
        \begin{equation} \lambda_- = k+l \quad \text{and} \quad \lambda_+ = k-l, \label{eqn:n=m_lambdas}\end{equation}
    respectively.
    
    \begin{proof}
        Since $G_A$ and $G_B$ are $k$-regular, by the Perron-Frobenius theorem,
            \[ A\,V_A=k\,V_A, \quad  B\,V_B=k\,V_B, \]
        and with $n=m$,
            \[ V_A = V_B = \onesqrtn\allonesn.\]
        Furthermore, by the $l$-regularity and choice of signs in $C$, one has
        \begin{equation}\label{eqn:l_reg_C_eigs}
            C\,V_B = -l\,V_A \quad \text{and} \quad C^T\,V_A = -l\,V_B .
        \end{equation}
        For
            \[ V_R^- = \onesqrttwo\begin{pmatrix} V_A \\[1mm] -V_B \end{pmatrix}, \]
        we compute:
        \begin{equation}\label{eqn:n=m_R_eigenvec}
            R\,V_R^- = \onesqrttwo\begin{pmatrix} A\,V_A - C\,V_B \\[1mm] C^T\,V_A - B\,V_B \end{pmatrix}
                     = \onesqrttwo\begin{pmatrix} k\,V_A + l\,V_A \\[1mm]  -l\,V_B - k\,V_B \end{pmatrix}
                     = (k+l)V_R^-.
        \end{equation}
        A similar calculation shows that
            \[ R\,V_R^+ = (k-l)V_R^+ .\]
    \end{proof}
\end{lemma}

The authors note that while the physical motivation from complex synchronized networks supports the use of $[C]_{ij}\in\{-1,0\}$, modifying \Cref{lem:basic_comp_eigvecs} to use $[C]_{ij}\in\{0, 1\}$ only changes the associated eigenvalues (switching the two emergent eigenvalues) of $V_R^-$ and $V_R^+$ from \Cref{eqn:n=m_lambdas} to
    \[ \lambda_- = k-l \quad \text{and} \quad \lambda_+ = k+l .\]

Furthermore, by requiring $V_R^-$ and $V_R^+$ to be eigenvectors of $R$, we have the following corollary.

\begin{corollary}[Order of Subgraphs of $R$]\label{cor:order_of_subgraphs}
    Let $G_R$ be constructed from two disjoint $k$-regular subgraphs $G_A$ and $G_B$, \underline{\emph{with orders $n, m$}} and entries $[A]_{ij}\,, [B]_{ij} \in\{0, 1\}$, coupled by a bipartite graph $G_C$ with undirected connection matrix $C$ (with $[C]_{ij}\in\{-1,0\}$).
    
    We have that $V_R^-$ and $V_R^+$ are normalized eigenvectors of $R$, as defined in \Cref{lem:basic_comp_eigvecs}, with dimensionality $n+m$ and eigenvalues $\lambda_-$ and $\lambda_+$, if and only if
    \begin{enumerate}[(a)]
        \item $n = m$, \label{cor:order:cond1}
        \item $C$ and $C^T$ are $l$-regular for some $l$ (each row and column of $C$ sums to $-l$), \label{cor:order:cond2}
        \item $\lambda_- = k+l$ and $\lambda_+ = k-l$ (for eigenvectors $V_R^-$ and $V_R^+$), and \label{cor:order:cond3}
        \item $R$ is $(k+l)$-regular. \label{cor:order:cond4}
    \end{enumerate}
    In particular, \cref{cor:order:cond2} is the necessary and sufficient condition and \cref{cor:order:cond3,cor:order:cond4} are sufficient. Furthermore, \ref{cor:order:cond4} $\iff$ \ref{cor:order:cond3} by Perron-Frobenius, \ref{cor:order:cond4} $\iff$ \ref{cor:order:cond2} by $k$-regularity of $A$ and $B$, and \ref{cor:order:cond2} $\implies$ $C$ is square ($nl = ml$) $\implies$ \ref{cor:order:cond1}. 
    \begin{proof}
        It suffices to show that $V_R^-$ and $V_R^+$ are eigenvectors of $R \iff$ \cref{cor:order:cond2}.
        \begin{itemize}
            \item[($\implies$)]
            For
                \[ V_R^+ = \onesqrttwo\begin{pmatrix} V_A \\[1mm] V_B \end{pmatrix}, \]
            we compute:
            \[
                R\,V_R^+ = \onesqrttwo\begin{pmatrix} A\,V_A + C\,V_B \\[1mm] C^T\,V_A + B\,V_B \end{pmatrix}
                         = \onesqrttwo\begin{pmatrix} k\,V_A + C\,V_B \\[1mm]   C\,V_A + k\,V_B \end{pmatrix}.
            \]
            By the definition of eigenvectors,
            \[
                \onesqrttwo\begin{pmatrix} k\,V_A + C\,V_B \\[1mm]  C\,V_A + k\,V_B \end{pmatrix}
                = \lambda_R \begin{pmatrix} V_A \\[1mm] V_B \end{pmatrix} ,
            \]
            which yields equations
            \begin{align}
                k\,V_A + C\,V_B = \lambda_R\,V_A \quad &\text{and} \quad C^T\,V_A + k\,V_B = \lambda_R\,V_B \nonumber\\
                \implies C\,V_B = (\lambda_R-k)\,V_A \quad &\text{and} \quad C^T\,V_A = (\lambda_R-k)\,V_B \,. \label{eqn:c_cT_forced_reg}
            \end{align}
            As $V_A = \allonesn$ and $V_B = \allonesm$, $C\,V_B$ and $C^T\,V_A$ must produce constants times all ones vectors, forcing all row sums to be equal and all column sums to be equal:
            \[ \sum_{i} [C]_{ij} = -l_A \quad \forall \,j \quad \text{and} \quad
               \sum_{j} [C]_{ij} = -l_B \quad \forall \,i\]
            Making \Cref{eqn:c_cT_forced_reg} into
            \begin{align*}
                -l_A\,V_A = (\lambda_R-k)\,V_A \quad &\text{and} \quad -l_B\,V_B = (\lambda_R-k)\,V_B \\
                \implies -l_A + k = \lambda_R \quad &\text{and} \quad -l_B + k = \lambda_R \\
                \implies & l_A = l_B
            \end{align*}

            Thus, $C$ and $C^T$ are $l$-regular.

            \item[($\impliedby$)]
            This follows from \ref{cor:order:cond2} $\iff$ \ref{cor:order:cond4} and either the Perron-Frobenius theorem or \ref{cor:order:cond1} and \Cref{lem:basic_comp_eigvecs}.
        \end{itemize}
    \end{proof}
\end{corollary}

\subsection{Arbitrary QL--Bit State}
Now, to construct an arbitrary QL--bit state, we allow the regularities of $A$ and $B$ to differ. Define tunable parameter
    \begin{equation}\label{eqn:delta_def}
        \Delta \coloneqq \frac{k_A-k_B}{2l}
    \end{equation}
as the average of the two subgraph eigenvalues divided by the regularity of the bipartite connection matrix $C$.

\begin{theorem}[Arbitrary Quantum-Like State Construction] \label{thm:arb_state_construction}
    Let $G_R$ be constructed from two disjoint regular subgraphs $G_A$ and $G_B$, with regularities $k_A,\, k_B$, each of order $n$, with entries $[A]_{ij}\,, [B]_{ij} \in\{0, 1\}$, coupled by an $l$-regular bipartite graph $G_C$ with undirected connection matrix $C$ (with $[C]_{ij}\in\{-1,0\}$). Let $V_A,\, V_B$ be the normalized eigenvectors corresponding to the largest eigenvalues $\lambda_A,\, \lambda_B$ of $A,\, B$, respectively, as in \Cref{eqn:vavb_eigvecs,eqn:vavb_eigvals}, and basis vectors $\kp,\,\km$ be defined as in \Cref{eqn:kp_km}. Let the vector
        \begin{gather}
            \kpsi \coloneqq V_R = a\,\kp+b\,\km = \begin{pmatrix} \omega_1 V_A \nonumber\\[1mm] \omega_2 V_B \end{pmatrix} \quad \text{with}\\
            \omega_1 \coloneqq \onesqrttwo (a+b),\quad \omega_2 \coloneqq \onesqrttwo (a-b), \quad s.t. \label{eqn:arb_qubit_def} \\
            |a|^2 + |b|^2 \equiv |\omega_1|^2 + |\omega_2|^2 = 1 \nonumber %
        \end{gather}
    define an arbitrary quantum-like state.
    For $|a| \neq |b|$, if the regularities $k_A,\,k_B,$ and $l$ are tuned proportionally by the condition
        \begin{equation}\label{eqn:delta_tuning_eqn}
            \Delta = \frac{2ab}{b^2-a^2} = \frac{\omega_2^2-\omega_1^2}{2\omega_1\omega_2} \,,
        \end{equation}
    then $\kpsi$ is an eigenvector of 
        \[ R=\begin{pmatrix} A & C \\[1mm] C^T & B \end{pmatrix} \]
    with eigenvalue
        \begin{equation}\label{eqn:arb_R_eigval}
            \lambda_R = k_A - \frac{\omega_2}{\omega_1}l \stackrel{(\ref{eqn:delta_def},\ref{eqn:delta_tuning_eqn})}{=} k_B - \frac{\omega_1}{\omega_2}l .
        \end{equation}
    (Notation $\stackrel{(\ref{eqn:delta_def},\ref{eqn:delta_tuning_eqn})}{=}$ indicates that step was taken using \Cref{eqn:delta_def,eqn:delta_tuning_eqn}.)

    \begin{proof}
        It suffices to show that $\kpsi$ is an eigenvector of $R$ with eigenvalue $\lambda_R$ given \Cref{eqn:delta_tuning_eqn}.
        
        Solving \Cref{eqn:delta_def,eqn:delta_tuning_eqn} for $k_A$ and $k_B$ yields
            \begin{equation} \label{eqn:arb_kakb_solvedfor}
                k_A = k_B + \frac{\omega_2^2 - \omega_1^2}{\omega_1\omega_2} l,\quad
                k_B = k_A - \frac{\omega_2^2 - \omega_1^2}{\omega_1\omega_2} l .
            \end{equation}
        We proceed with solving the eigenvalue equation.
        \begin{align}
            R\kpsi &= \begin{pmatrix} A & C \\[1mm] C^T & B \end{pmatrix} \begin{pmatrix} \omega_1 V_A \\[1mm] \omega_2 V_B \end{pmatrix}\\
                   &= \begin{pmatrix} \omega_1 AV_A   + \omega_2 CV_B \\[1mm]
                                      \omega_1 C^TV_A + \omega_2 BV_B \end{pmatrix}\\
                   &\stackrel{(\ref{eqn:vavb_eigvals},\ref{eqn:l_reg_C_eigs})}{=}
                    \begin{pmatrix} \omega_1 k_A V_A - \omega_2 l V_A \\[1mm]
                                  - \omega_1 l V_B   + \omega_2 k_B V_B \end{pmatrix}\\
                   &\stackrel{(\ref{eqn:arb_kakb_solvedfor})}{=}
                    \begin{pmatrix}
                        \left(\omega_1 \frac{\omega_2^2 - \omega_1^2}{\omega_1\omega_2} l + \omega_1 k_B - \omega_2 l \right) V_A \\[2mm]
                        \left(-\omega_2 \frac{\omega_2^2 - \omega_1^2}{\omega_1\omega_2} l + \omega_2 k_A - \omega_1 l \right) V_B
                    \end{pmatrix}\\
                   &= \begin{pmatrix}
                        \left(-\frac{\omega_1^2}{\omega_2} l + \omega_1 k_B \right) V_A \\[1mm]
                        \left(\frac{-\omega_2^2}{\omega_1} l + \omega_2 k_A \right) V_B
                    \end{pmatrix}\\
                   &\stackrel{(\ref{eqn:arb_R_eigval})}{=}
                   \begin{pmatrix} \omega_1\lambda_R V_A \\[1mm] \omega_2\lambda_R V_B \end{pmatrix}
                    = \lambda_R \begin{pmatrix} \omega_1 V_A \\[1mm] \omega_2 V_B \end{pmatrix}
        \end{align}
        (Once more for cross-field notation clarity: $\stackrel{(\ref{eqn:vavb_eigvals},\ref{eqn:l_reg_C_eigs})}{=}$ indicates that step was taken using \Cref{eqn:vavb_eigvals,eqn:l_reg_C_eigs}.)
    \end{proof}
\end{theorem}

In the same notion as after \Cref{lem:basic_comp_eigvecs}, using $[C]_{ij}\in\{0, +1\}$ only changes \Cref{eqn:arb_R_eigval} to 
    \begin{equation}\label{eqn:arb_R_eigval_C_pos}
        \lambda_R = k_A + \frac{\omega_2}{\omega_1}l = k_B + \frac{\omega_1}{\omega_2}l .
    \end{equation}
The math in \Cref{thm:arb_state_construction} remains the same except that $l$ flips signs in every equation, including in the definition of $\Delta$ (\Cref{eqn:delta_def}).

Thus, any state of the form $a\,\psi_{+}+b\,\psi_{-}$ (with $a^2+b^2=1$) may be realized via appropriate choices of $k_A$, $k_B$, and $l$ (i.e. $\Delta$), provided $|a| \neq |b|$. However, note that as $|a| \to |b| = \onesqrttwo$ we have
\begin{equation}\label{eqn:delta_limit}
    \lim_{|a| \to |b|}\Delta=\infty.
\end{equation}
This divergence implies that a perfectly balanced state via detuning alone would require an unbounded difference in the subgraph regularities.  This is further illustrated in \Cref{fig:basic_visualization}, where the set of possible $\kpsi$ values (cylinder, defined by $|a|^2+|b|^2=1$ for $a, b \equiv x-$axis, $y-$axis), plotted against tunable parameter $\Delta \equiv z-$axis, are constrained to the feasible regions (lines running vertically along the walls of the cylinder) of \Cref{eqn:delta_tuning_eqn}.  Indeed, as $|x| \to |y|$, the feasible regions grow to $\infty$.  A solution to this is found in the proof for the asymmetric coupling construction, \Cref{lem:asym_coup_eigvecs} (by inverting its $\Delta_C$), and applied to symmetric coupling in \Cref{sec:asym:switching_sym}.

\subsection{Constraints on Symmetric Coupling}\label{sec:symm:constraints}

We proceed with an explicit construction of $\Delta$ given a target qubit state $\kpsi$.

\begin{example}[Constructing a Skewed Superposition via Detuning]
    Given a target single-qubit state $\kpsi$ determined by parameters $a = \sqrt{\frac{1}{3}},\, b=\sqrt{\frac{2}{3}}$, we can construct a graph based on the following method:
    \begin{enumerate}
        \item Compute $\Delta$ as $\Delta = \frac{2ab}{b^2-a^2} = \frac{2\sqrt{2/9}}{1/3} = 2\sqrt{2} \approx 2.8284$
        \item Set the regularities of $A,B,$ and $C$ such that $2\sqrt{2} = \frac{k_A-k_B}{2l}$.  For example, in a $n/2=30$ node graph (each subgraph having 30 nodes), if $k_A$ and $k_B$ are set prior, in order to get a more robust emergent state, as $k_A = 25,\, k_B = 20$, then $l = \frac{5\sqrt{2}}{8} \approx 0.8838835$.
    \end{enumerate}
\end{example}

One may notice that this example sets the regularity of the bipartite connection matrix $C$ to a non-integer value.  While mathematically this matters little if we allow $[C]_{ij} \in \R$ (or even $\in \C$), as discussed in \Cref{sec:real_valued_C}, using the graph intuition, where regularity is defined by the number of edges coming out of a particular node, non-integer values do not make sense.  This constraint is tightened further by restricting to simple graphs (no multiedges nor self-loops).  In particular, for a graph $G_R$ with $2n$ nodes ($n$ nodes per subgraph $G_A$ and $G_B$), one can impose the constraints
\begin{align}\label{eqn:delta_int_constraints}
    |k_A - k_B| < n,& \nonumber\\
    |l| < n,& \nonumber\\
    k_A, k_B, l \neq 0,& \quad \text{and}\\
    k_A, k_B, l \in \Z.& \nonumber
\end{align}
Which, intuitively, say that the valency of any one node cannot be more than the number of nodes in the target subgraph (first and second conditions), that no graph can have a regularity of 0 (no edges; third condition), and that the regularities of all three graphs (including bipartite connection graph $G_C$) must to be integers (last condition).  Together, these give the exact bound
    \[ |\Delta| < n,\quad \Delta \in \Q \]
(for the group of rational numbers $\Q$) in the maximal setting of $k_A = k_B-1$ and $l=1$ (see \Cref{eqn:delta_def}).  Furthermore, we can precisely relate the possible values of $a$ and $b$ in \Cref{eqn:arb_qubit_def} to $\Delta$ by the bounds
\begin{align}\label{eqn:ab_dfn_by_delta}
    a &= \pm \onesqrttwo\sqrt{\left( 1\pm \frac{1}{\sqrt{\Delta^2 + 1}}\right)}\nonumber\\
    b &= \pm \onesqrttwo\sqrt{\left( 1\mp \frac{1}{\sqrt{\Delta^2 + 1}}\right)}
\end{align}
given $a^2+b^2=1 \implies b = \pm \sqrt{1-a^2}$.  This leads to 4 possible sign combinations indicating what values of $a$ and $b$ each formula is valid for,
\begin{align}\label{eqn:ab_sign_combos_delta}
    a(+, +) \implies +a, |a| > |b| \nonumber\\
    a(+, -) \implies +a, |b| > |a| \nonumber\\
    a(-, +) \implies -a, |a| > |b| \\
    a(-, -) \implies -a, |b| > |a| \nonumber
\end{align}
where $a(\cdot, \cdot)$ informally references a choice of signs in \Cref{eqn:ab_dfn_by_delta}.  Perhaps more intuitively, these sign combinations define 4 quadrants (extended to 3 dimensions by $\Delta$) of $a$ vs $b$ defined by boundaries that go to $\infty$ as $|a| \to |b|$.

\begin{figure}
    \centering
    \includegraphics[width=0.5\linewidth]{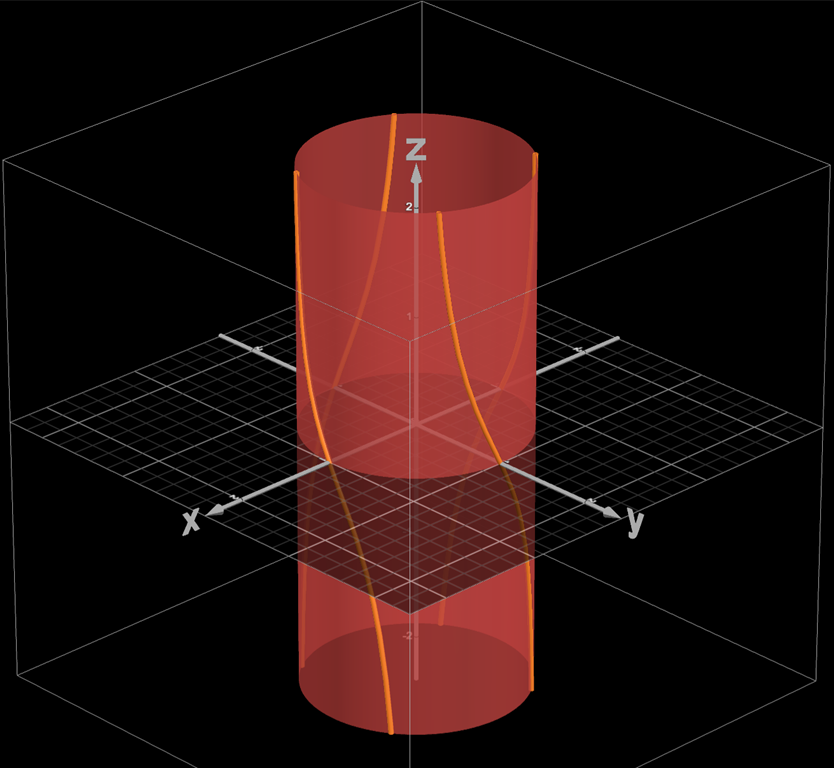}
    \caption{Basic visualization of the bounds of $\Delta$ applied to $\kpsi$.  The x-axis is $a$, the y-axis is $b$, and the z-axis is $\Delta$.  The cylinder is defined by the circle (extended to 3D) $a^2+b^2=1$ and is the set of possible points under the definition of $\kpsi$ in \Cref{eqn:arb_qubit_def}.  By plotting it against $\Delta$, we can see what values $\kpsi$ are possible given the equation for $\Delta$, \Cref{eqn:delta_tuning_eqn}, as shown by the four lines running up the cylinder.  See text after \Cref{thm:arb_state_construction} for more.}
    \label{fig:basic_visualization}
\end{figure}

\begin{figure}
    \centering
    \includegraphics[width=0.5\linewidth]{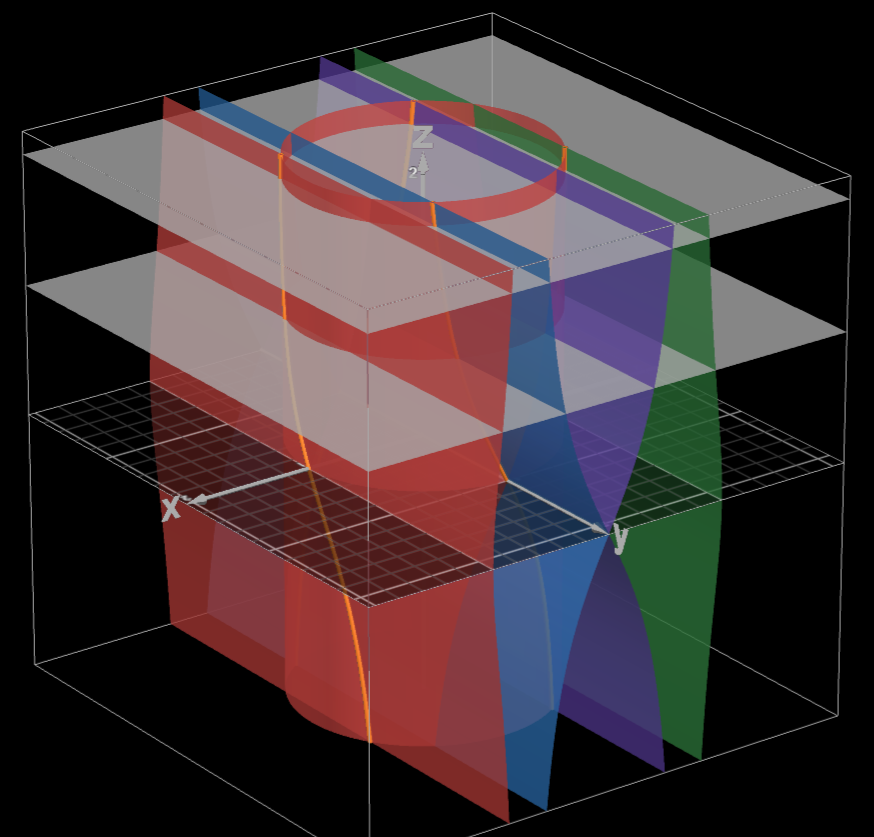}
    \caption{Full visualization of the bounds of $\Delta$ applied to $\kpsi$.  The x-axis is $a$, the y-axis is $b$, and the z-axis is $\Delta$. Refer to \Cref{fig:basic_visualization} for introduction.  Added to this plot are six planes.  The vertical four planes are defined by the four variations of $a$ in \Cref{eqn:ab_dfn_by_delta} ($\pm \otimes \pm$) and demonstrate the possible $a$ values given tunable parameter $\Delta$.  As analyzed in \Cref{eqn:delta_int_constraints}, $\Delta$ is also constrained to be a rational number. For simplicity, the plot shows only positive integer values of $\Delta$ as xy-planes perpendicular to the z-axis.  Feasible values for $\kpsi$ are therefore constrained to intersections of the xy-planes with the feasible vertical lines.}
    \label{fig:full_visualization}
\end{figure}

These practical constraints on $\Delta$ are illustrated in \Cref{fig:full_visualization}.  The constraint $\Delta \in \Q$ is shown by the xy-planes at (demonstratively) integer values on the z-axis ($\equiv \Delta$).  The four possible $a$ combinations ($\pm \otimes \pm$) in \Cref{eqn:ab_dfn_by_delta} are shown as vertical planes and intersect precisely with the feasible lines more easily seen in \Cref{fig:basic_visualization} and discussed above.  The possible values for $\kpsi$ are therefore understood as the intersection of the xy-planes with the feasible lines.  As $n$ increases, the ability of \Cref{eqn:delta_def} to approximate $\kpsi$ increases precisely as the growth of a denominator increases the ability of a fraction to approximate a real number (i.e. the rational numbers).
We provide a list of equations to reconstruct the graph in \Cref{app:fancy_plot_eqns}.

\section{Asymmetric Coupling Construction}\label{sec:asym}
To avoid the divergence in $\Delta$ (when $|a|\to |b| \equiv \omega_1,\,\omega_2 \to 0$), we allow the coupling to be directed. Replace the single matrix $C$ by two matrices $C_A$ and $C_B$:
    \[ R = \begin{pmatrix} A & C_A \\[1mm] C_B & B \end{pmatrix}, \]
with
    \begin{equation}\label{eqn:lab_reg_Cab_eigs}
        C_A\,V_B = -l_A\,V_A,\quad C_B\,V_A = -l_B\,V_B.
    \end{equation}
By defining the tunable parameter
    \begin{equation}\label{eqn:delta_c_def}
        \Delta_C \coloneqq \frac{l_A}{l_B},
    \end{equation} 
and sets $l_A = 0$, one obtains the emergent eigenvector
    \[ V_R = \begin{pmatrix} 0 \\[1mm] V_B \end{pmatrix} = \onesqrttwo(\kp-\km) = \ko, \]
one of the two states of issue in \Cref{eqn:delta_limit} ($a\to -b = -\onesqrttwo, \ko$).  The other ($a \to b = \onesqrttwo, \kz$) is created when $l_B = 0$. More generally, by tuning $l_A$ and $l_B$ independently one can realize any desired (near-balanced) state without forcing $\Delta_C$ to diverge.

\begin{definition}[Directed Regular Graph]\label{dfn:directed_regular_graphs}
    An unweighted directed graph $G_D$ is $l$--regular if every vertex has out-degree $l$, equivalently,
        \[ \sum_{j=1}^n [D]_{ij}=l \quad \forall\,i . \]
    For the signed connection matrices $C_A$ and $C_B$ used below, $l_A$ and $l_B$ denote the corresponding unsigned out-degrees, so the rows of $C_A$ and $C_B$ sum to $-l_A$ and $-l_B$, respectively.
\end{definition}

\begin{lemma}[Balanced QL--Bit via Asymmetric Coupling]\label{lem:asym_coup_eigvecs}
    Let $G_R$ be constructed from two disjoint regular subgraphs $G_A$ and $G_B$, with regularities $k_A,\, k_B$, each of order $n$, with entries $[A]_{ij}\,, [B]_{ij} \in\{0, 1\}$, coupled by bipartite graph $G_C$ with $l_A,\,l_B$-regular \emph{directed} connection matrices $C_A, C_B$ (with $[C_A]_{ij},\,[C_B]_{ij}\in\{-1,0\}$). Let $V_A,\, V_B$ be the normalized eigenvectors corresponding to the largest eigenvalues $\lambda_A,\, \lambda_B$ of $A,\, B$, respectively, as in \Cref{eqn:vavb_eigvecs,eqn:vavb_eigvals}, and basis vectors $\kp,\,\km$ be defined as in \Cref{eqn:kp_km}. Let the vector $\kpsi$ be an arbitrary quantum-like state with parameters $a, b, \omega_1, \omega_2$ be defined as in \Cref{eqn:arb_qubit_def}.  If the regularities of the subgraphs are equal, $k_A = k_B$, and $l_A,\, l_B$ are tuned proportionally by the condition
    \begin{equation}\label{eqn:delta_c_tuning_eqn}
        \Delta_C = \frac{(a+b)^2}{(a-b)^2} = \frac{\omega_1^2}{\omega_2^2}
    \end{equation}
    then $\kpsi$ is an eigenvector of 
        \[ R=\begin{pmatrix} A & C_A \\[1mm] C_B & B \end{pmatrix} \]
    with eigenvalue
        \begin{align} \begin{split} \label{eqn:arb_R_eigval_directed_c}
            \lambda_R &= k_A - \frac{\omega_2}{\omega_1}l_A \stackrel{(\ref{eqn:delta_c_def},\ref{eqn:delta_c_tuning_eqn})}{=} k_B - \frac{\omega_1}{\omega_2}l_B\\
                      &= k - \frac{\omega_2}{\omega_1}l_A \;\;\stackrel{(\ref{eqn:delta_c_def},\ref{eqn:delta_c_tuning_eqn})}{=} k - \frac{\omega_1}{\omega_2}l_B
        \end{split} \end{align}
    similarly to \Cref{thm:arb_state_construction}, \Cref{eqn:arb_R_eigval}.

    \begin{proof}
        We note for completeness that by definition (\Cref{eqn:arb_qubit_def}), it cannot be the case that $\omega_1 = \omega_2 = 0$.  Thus, if $\omega_1 \text{ XOR } \omega_2 = 0$, one may simply invert the definition and target equation for $\Delta_C$ to achieve the desired state, as the equations derived for $\Delta_C$ began as $\omega_1^2\,l_B-\omega_2^2\,l_A = 0$.
        Further details provided in \Cref{app:general_characteristic_eqn}.

        Otherwise, it suffices to show that $\kpsi$ is an eigenvector of $R$ with eigenvalue $\lambda_R$ given \Cref{eqn:delta_c_tuning_eqn}.

        Solving \Cref{eqn:delta_c_def,eqn:delta_c_tuning_eqn} for $l_A$ and $l_B$ yields
            \begin{equation} \label{eqn:arb_lalb_solvedfor}
                l_A = \frac{\omega_1^2}{\omega_2^2} l_B,\quad
                l_B = \frac{\omega_2^2}{\omega_1^2} l_A .
            \end{equation}
        We proceed with solving the eigenvalue equation.
        \begin{align}
            R\kpsi &= \begin{pmatrix} A & C_A \\[1mm] C_B & B \end{pmatrix} \begin{pmatrix} \omega_1 V_A \\[1mm] \omega_2 V_B \end{pmatrix}\\
                   &= \begin{pmatrix} \omega_1 AV_A   + \omega_2 C_AV_B \\[1mm]
                                      \omega_1 C_BV_A + \omega_2 BV_B \end{pmatrix}\\
                   &\stackrel{(\ref{eqn:vavb_eigvals},\ref{eqn:lab_reg_Cab_eigs})}{=}
                    \begin{pmatrix} \omega_1 k V_A - \omega_2 l_A V_A \\[1mm]
                                  - \omega_1 l_B V_B   + \omega_2 k V_B \end{pmatrix}\\
                   &= \begin{pmatrix}
                        \omega_1 \left( k - \frac{\omega_2}{\omega_1} l_A \right) V_A \\[1mm]
                        \omega_2 \left( k - \frac{\omega_1}{\omega_2} l_B \right) V_B
                    \end{pmatrix}\\
                   &\stackrel{(\ref{eqn:arb_R_eigval_directed_c})}{=}
                   \begin{pmatrix} \omega_1\lambda_R V_A \\[1mm] \omega_2\lambda_R V_B \end{pmatrix}
                    = \lambda_R \begin{pmatrix} \omega_1 V_A \\[1mm] \omega_2 V_B \end{pmatrix}
        \end{align}
    \end{proof}
\end{lemma}

Explicitly, the primary differences between \Cref{lem:asym_coup_eigvecs} and \Cref{thm:arb_state_construction} are the following:
\begin{enumerate}
    \item The lemma allows $C_A \neq C_B^T$, i.e. the bipartite connection graph $G_C$ to be directed, but still require that both be regular (see \Cref{dfn:directed_regular_graphs}) with entries $\in [-1, 0]$.  This loosens the setting.
    \begin{itemize}
        \item The interpretation of $C_A$ is all of the outgoing edges from $A \to B$, and its (row) regularity is over the outgoing edges from $A$.  Similarly for $C_B$.
    \end{itemize}

    \item The lemma tightens the setting by requiring $k_A = k_B$, the regularities of $A$ and $B$ to match.
    The connection between these two relaxations is made explicit in \Cref{app:general_characteristic_eqn}.

    \item The theorem requires $|a|\neq |b| \equiv (\omega_1 \text{ or } \omega_2) \neq 0$, but the lemma has no such constraint. If $(a=b \text{ or } a=-b) \equiv (\omega_1 \text{ or } \omega_2) = 0$, one may simply invert the definition and target equation for $\Delta_C$ to achieve the desired state.  Both $a$ and $b$ cannot be simultaneously 0 by definition.

    \item $\Delta_C >= 0$: since $[C]_{ij}$ are all the same sign (or 0), $\Delta_C \coloneqq \frac{l_A}{l_B} >=0$.  $\Delta$ has $k_A - k_B$, where trivially swapping the larger magnitude flips the sign of $\Delta$.
\end{enumerate}

\noindent In the same notion as after \cref{lem:basic_comp_eigvecs,thm:arb_state_construction}, using $[C]_{ij}\in\{0, +1\}$ only changes \Cref{eqn:arb_R_eigval_directed_c} to 
    \begin{equation}\label{eqn:arb_R_eigval_directed_C_pos}
        \lambda_R = k + \frac{\omega_2}{\omega_1}l_A = k + \frac{\omega_1}{\omega_2}l_B .
    \end{equation}
The math in \Cref{lem:asym_coup_eigvecs} remains the same except that $l_A, l_B$ flip signs in every equation.

Thus, any state of the form $a\,\psi_{+}+b\,\psi_{-}$ (with $a^2+b^2=1$) may be realized via appropriate choices of $l_A$ and $l_B$ (i.e. $\Delta_C$).  If $l_A = 0$ or $l_B = 0$, it is trivial to define $\Delta_C = \Delta_C\inv$ and invert \Cref{eqn:delta_c_tuning_eqn}.  As discussed in the proof, by definition, it cannot be the case that $\omega_1 = \omega_2 = 0$.

This divergence implies that a perfectly balanced state via detuning alone would require an unbounded difference in the subgraph regularities.  This is further illustrated in \Cref{fig:basic_visualization}, where the set of possible $\kpsi$ values (cylinder, defined by $a, b \equiv x-$axis, $y-$axis) are constrained to the feasible regions (lines running vertically along the walls of the cylinder) of \Cref{eqn:delta_c_tuning_eqn}.  Indeed, as $|x| \to |y|$, the feasible regions grow to $\infty$.

\subsection{Constraints on Asymmetric Coupling}\label{sec:asymm:constraints}
We proceed with an explicit construction of $\Delta$ given a target qubit state $\kpsi$.

\begin{example}[Constructing a Skewed Superposition via Detuning]
    Given a target single-qubit state $\kpsi$ determined by parameters $a = \sqrt{\frac{1}{3}},\, b=\sqrt{\frac{2}{3}}$, we can construct a graph based on the following method:
    \begin{enumerate}
        \item Compute $\Delta_C$ as $\Delta_C = \left(\frac{a+b}{a-b}\right)^2 = 17+12\sqrt2 \approx 33.9706$

        \item Set the regularities of $A,B,$ and $C$ such that $17+12\sqrt2 = \frac{l_A}{l_B}$.  For example, in a $n/2=150$ node graph (each subgraph having $150$ nodes), if we set a minimum threshold for the regularities $l_A,\,l_B$ to $3$ in order to get a more robust emergent state, then $l_B = 3,\; l_A = 3*(17+12\sqrt2)\approx 101.91$.
    \end{enumerate}
\end{example}

One may notice that this example sets the regularity of the bipartite connection matrix $C$ to a non-integer value.  While mathematically this matters little if we allow $[C]_{ij} \in \R$ (or even $\in \C$), as discussed in \Cref{sec:real_valued_C}, using the graph intuition, where regularity is defined by the number of edges coming out of a particular node, non-integer values do not make sense.  This constraint is tightened further by restricting to simple graphs (no multiedges nor self-loops).  In particular, for a graph $G_R$ with $2n$ nodes ($n$ nodes per subgraph $G_A$ and $G_B$), one can impose the constraints
\begin{align}\label{eqn:delta_c_int_constraints}
    |l_A - l_B| < n,& \nonumber\\
    l_A, l_B, k \neq 0,& \quad \text{and}\\
    l_A, l_B, k \in \Z.& \nonumber
\end{align}
Which, intuitively, say that the valency of any one node cannot be more than the number of nodes in the target subgraph (first condition), that no graph can have a regularity of 0 (no edges; second condition), and that the regularities of all three graphs must to be integers (last condition).  Together, these give the exact bound
    \[ |\Delta_C| < n,\quad \Delta_C \in \Q \]
(for the group of rational numbers $\Q$) in the maximal setting of $l_A = n-1$ and $l_B=1$ (see \Cref{eqn:delta_c_def}).  Furthermore, we can precisely relate the possible values of $a$ and $b$ in \Cref{eqn:arb_qubit_def} to $\Delta_C$
by the bounds
\begin{align}\label{eqn:ab_dfn_by_delta_c}
    ab &= \frac{\Delta_C-1}{2(\Delta_C+1)} \nonumber \implies\\ 
    a &= \pm \sqrt{\frac12 \pm \frac{\sqrt{\Delta_C}}{\Delta_C + 1}}\nonumber\\
    b &= \pm \sqrt{\frac12 \pm \frac{\sqrt{\Delta_C}}{\Delta_C + 1}} \; ,
\end{align}

given $a^2+b^2=1 \implies b = \pm \sqrt{1-a^2}$.  This leads to 4 possible sign combinations indicating what values of $a$ and $b$ each formula is valid for,
\begin{align}\label{eqn:ab_sign_combos_delta_c}
    a(+, +) \implies +a, |a| > |b| \nonumber\\
    a(+, -) \implies +a, |b| > |a| \nonumber\\
    a(-, +) \implies -a, |a| > |b| \\
    a(-, -) \implies -a, |b| > |a| \nonumber
\end{align}
where $a(\cdot, \cdot)$ informally references a choice of signs in \Cref{eqn:ab_dfn_by_delta_c}.  Perhaps more intuitively, as before, these sign combinations define 4 quadrants (extended to 3 dimensions by $\Delta_C$) of $a$ vs $b$ defined by boundaries that go to $\infty$ as $|a| \to |b|$.  Notably, there is no difference between \Cref{eqn:ab_sign_combos_delta_c} and \Cref{eqn:ab_sign_combos_delta}.

One can follow a similar process for $\Delta_C\inv$ using the relation
\begin{equation}\label{eqn:ab_dfn_by_delta_c_inv}
    ab = - \frac{\Delta_C\inv-1}{2(\Delta_C\inv+1)}
\end{equation}
where the only difference is the first sign.

\begin{figure}
    \centering
    \includegraphics[width=0.5\linewidth]{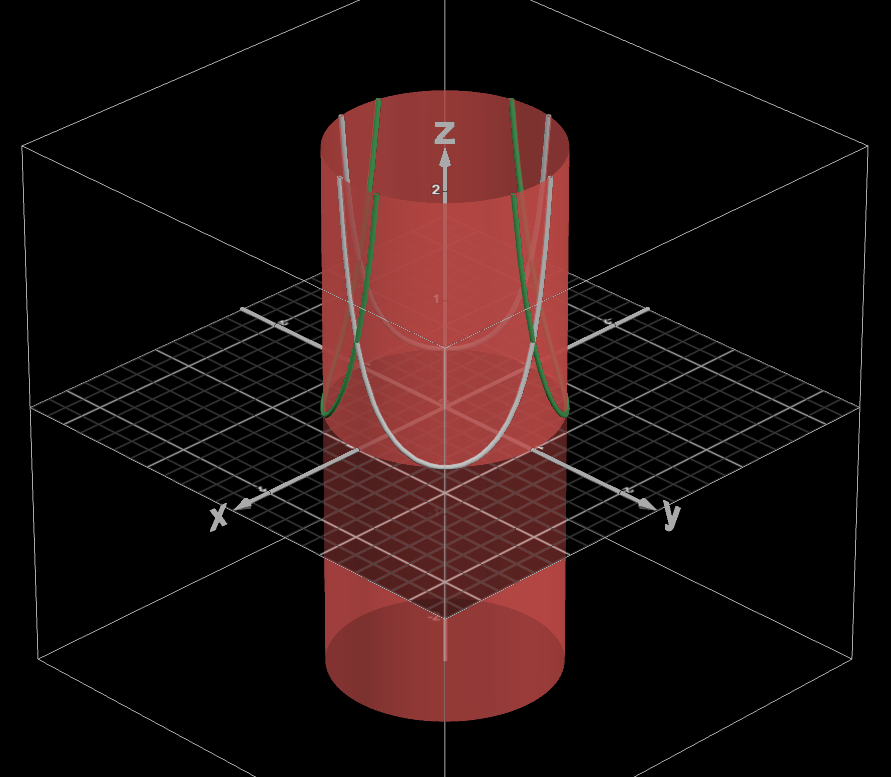}
    \caption{Basic visualization of the feasible values of $\Delta_C$ (green line, $\to \infty$ as $x\to y$) and $\Delta_C\inv$ (grey line, $\to \infty$ as $x \to -y$) applied to $\kpsi$.  The x-axis is $a$, the y-axis is $b$, and the z-axis is $\Delta_C \; (\Delta_C\inv)$.  The cylinder is defined by the circle (extended to 3D) $a^2+b^2=1$ and is the set of possible points under the definition of $\kpsi$ in \Cref{eqn:arb_qubit_def}.  By plotting it against $\Delta_C \; (\Delta_C\inv)$, we can see what values $\kpsi$ are possible given the equation for $\Delta_C \; (\Delta_C\inv)$, \Cref{eqn:delta_c_tuning_eqn}, as shown by the four lines running up the cylinder.  See text after \Cref{lem:asym_coup_eigvecs} for more.}
    \label{fig:basic_visualization_delta_c}
\end{figure}

\begin{figure}[ht]
    \centering
    \includegraphics[width=0.5\linewidth]{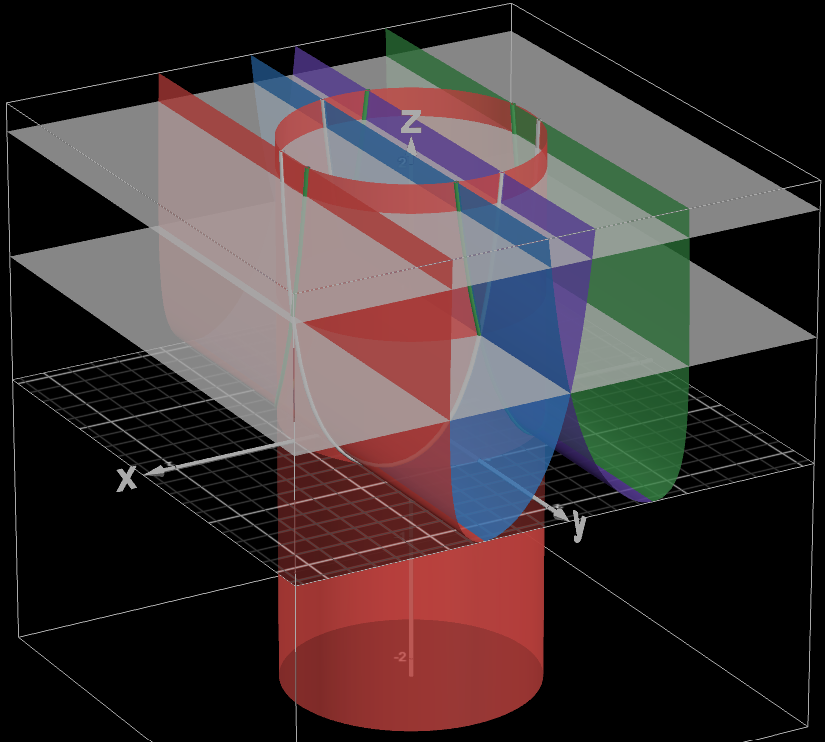}
    \caption{Full visualization of the feasible values of $\Delta_C$ (green line, $\to \infty$ as $x\to y$) and $\Delta_C\inv$ (grey line, $\to \infty$ as $x \to -y$) applied to $\kpsi$.  The x-axis is $a$, the y-axis is $b$, and the z-axis is $\Delta$. Refer to \Cref{fig:basic_visualization_delta_c} for introduction.  Added to this plot are six planes.  The four vertical planes are defined by the four variations of $a$ in \Cref{eqn:ab_dfn_by_delta_c} ($\pm \otimes \pm$) and demonstrate the possible $a$ values given tunable parameter $\Delta_C \; (\Delta_C\inv)$.  As analyzed in \Cref{eqn:delta_c_int_constraints}, $\Delta_C$ is also constrained to be a rational number. For simplicity, the plot shows only positive integer values of $\Delta_C$ as xy-planes perpendicular to the z-axis.  Feasible values for $\kpsi$ are therefore constrained to intersections of the xy-planes with the feasible vertical lines.}
    \label{fig:full_visualization_delta_c}
\end{figure}

These practical constraints on $\Delta_C \; (\Delta_C\inv)$ are illustrated in \Cref{fig:full_visualization_delta_c}.  The constraint $\Delta_C \in \Q$ is shown by the xy-planes at (demonstratively) integer values on the z-axis ($\equiv \Delta_C, \Delta_C\inv$).  The two possible $a$ combinations ($\pm$) in \Cref{eqn:ab_dfn_by_delta_c} are shown as vertical planes and intersect precisely with the feasible lines more easily seen in \Cref{fig:basic_visualization_delta_c} and discussed above.  The possible values for $\kpsi$ are therefore understood as the intersection of the xy-planes with the feasible lines.  As $n$ increases, the ability of \Cref{eqn:delta_c_def} to approximate $\kpsi$ increases precisely as the growth of a denominator increases the ability of a fraction to approximate a real number (i.e. the rational numbers).
We provide a list of equations to reconstruct the graph in \Cref{app:fancy_plot_eqns}.

\subsection{Switching Between Asymmetric Constructions}\label{sec:asym:switching_asym}
With simple observation of \Cref{fig:basic_visualization_delta_c,fig:full_visualization_delta_c} and \Cref{eqn:ab_sign_combos_delta_c}, we provide the following simple switching condition on which of $\Delta_C$ or $\Delta_C\inv$ to choose:

\begin{equation}\label{eqn:switching_condition_asym} %
    \begin{cases}
        \Delta_C \quad \; \, \text{if } \sign(a) = - \sign(b)\\
        \Delta_C\inv \quad \text{if } \sign(a) = \sign(b)
    \end{cases}
\end{equation}

And $\Delta_C = \Delta_C\inv$ when $a=0$ or $b=0$.

\subsection{Switching Between Symmetric Constructions}\label{sec:asym:switching_sym}
Similarly to asymmetric coupling, we can invert $\Delta$ in \Cref{eqn:delta_def,eqn:delta_tuning_eqn} to completely cover the divergence in \Cref{eqn:delta_limit}.  This leads to an altered relationship between $\Delta\inv$ and $a, b$ (compare: \Cref{eqn:ab_dfn_by_delta}) defined by 
\begin{align}\label{eqn:ab_dfn_by_delta_inv}
    a &= \pm \onesqrttwo\sqrt{\left( 1\pm \frac{\Delta}{\sqrt{\Delta^2 + 1}}\right)}\nonumber\\
    b &= \pm \onesqrttwo\sqrt{\left( 1\mp \frac{\Delta}{\sqrt{\Delta^2 + 1}}\right)}
\end{align}
and possible sign combinations (indicating what values of $a$ and $b$ each formula is valid for),
\begin{align}\label{eqn:ab_sign_combos_delta_inv}
    a(+, +) \implies +a, -b \nonumber\\
    a(+, -) \implies +a, +b \nonumber\\
    a(-, -) \implies -a, +b \\
    a(-, +) \implies -a, -b \nonumber
\end{align}
where $a(\cdot, \cdot)$ informally references a choice of signs in \Cref{eqn:ab_dfn_by_delta_inv}.

\begin{figure}
    \centering
    \includegraphics[width=0.5\linewidth]{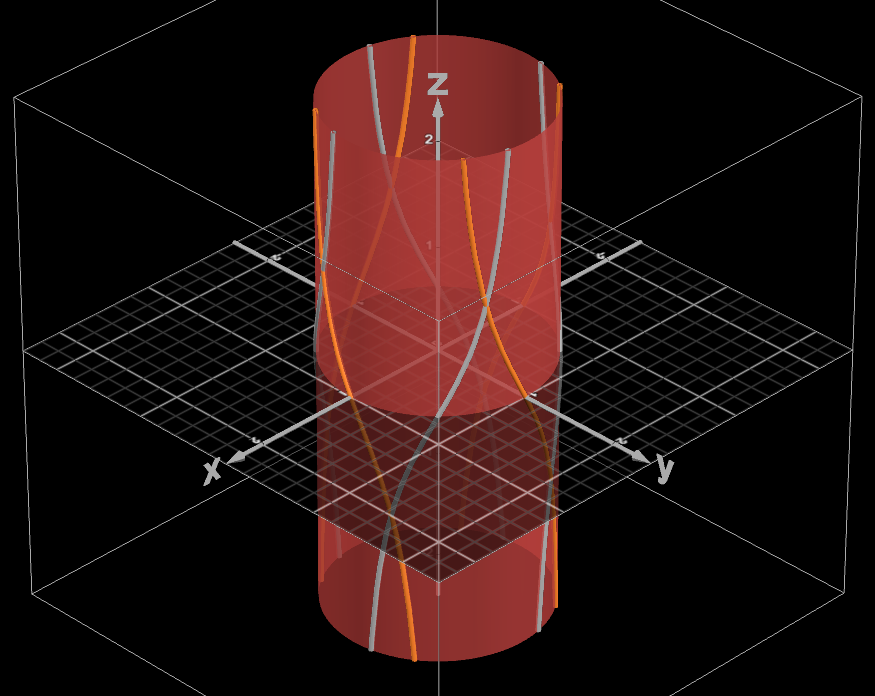}
    \caption{Symmetric Coupling: Basic visualization of the feasible values of $\Delta$ (orange line, $\to \infty$ as $|x| \to |y|$) and $\Delta\inv$ (grey line, $\to \infty$ as $x, y\to 0$) applied to $\kpsi$.  The x-axis is $a$, the y-axis is $b$, and the z-axis is $\Delta \; (\Delta\inv)$.  The cylinder is defined by the circle (extended to 3D) $a^2+b^2=1$ and is the set of possible points under the definition of $\kpsi$ in \Cref{eqn:arb_qubit_def}.  By plotting it against $\Delta \; (\Delta\inv)$, we can see what values $\kpsi$ are possible given the equation for $\Delta \; (\Delta\inv)$, \Cref{eqn:delta_tuning_eqn}, as shown by the eight lines running up the cylinder.  See \Cref{sec:asym:switching_sym} for more.}
    \label{fig:basic_visualization_delta_delta_inv}
\end{figure}

The feasible lines for $\Delta$ and $\Delta\inv$ can be seen together in \Cref{fig:basic_visualization_delta_delta_inv}.  Full visualization with \Cref{eqn:ab_dfn_by_delta_inv}, as in \Cref{fig:full_visualization},
is omitted for clarity but is available for reconstruction in \Cref{app:fancy_plot_eqns}.

To determine when to switch between $\Delta$ and $\Delta\inv$, we set $\Delta = \Delta\inv$ ($= \pm 1$) to get
    \[ a = \pm b (1 \pm \sqrt{2}) .\]
Plugging in $a^2+b^2=1$,
    \begin{equation*}
        a = \pm \sqrt{\frac{2 \pm \sqrt{2}}{4}} \quad \text{and} \quad 
        b = \pm \sqrt{\frac{2 \mp \sqrt{2}}{4}},
    \end{equation*}
agreeing with the visible intersection points in \Cref{fig:basic_visualization_delta_delta_inv}.  We therefore provide the following switching condition on which of $\Delta$ or $\Delta\inv$ to choose:

\begin{equation}\label{eqn:switching_condition_sym}
    \begin{cases}
        \Delta \quad \;\;\; \, \text{if } |a| \text{ or } |b| \ge \frac12\sqrt{2+\sqrt{2}} \\%
        \Delta\inv \quad \text{otherwise}
    \end{cases}
\end{equation}

And $\Delta = \Delta\inv$ when $|a|=$ or $|b|=\frac12\sqrt{2+\sqrt{2}}$.

\section{Random Walk Interpretation}\label{sec:random_walks}

As an alternative to the quantum‐state picture, we can view the composite graph $G_R$ through the lens of classical random walks.  Classically, the steady state (or stationary distribution) for a random walk on a graph is a vector containing the probability of being in each node if one ``wanders'' from node to node uniformly randomly along edges.  See \cite{lovasz1993random} for background on random walks. By tracing how the structural parameters $(k_A,k_B,l)$ fix both the target quantum‐like state $\kpsi=a\kz+b\ko$ and the walk’s stationary distribution, we see which degrees of freedom remain to optimize secondary criteria.

\subsection{Fully Regular Graph}
Let $G_R$ be a (strongly) connected $k$-regular graph on $2n$ vertices, with adjacency matrix
    \[ R = \begin{pmatrix} A & C\\ C^T & B \end{pmatrix}, \qquad \deg(i)=k\quad\forall i. \]
Define the simple random walk transition matrix
    \[ P=\frac1k R .\]
Since $G_R$ is $k$-regular, $P$ is doubly‐stochastic (each row and column is a probability vector over next steps) and, by connectivity, irreducible. Thus, there is a unique stationary distribution, namely the uniform one.  In the classical (probabilistic) picture we normalize in the $L^1$‐sense:
    \[ \pi = \left( \frac1{2n},\frac1{2n},\dots,\frac1{2n} \right), \quad \pi\,P = \pi, \quad \sum_i\pi_i=1 . \]
Equivalently, by Perron–Frobenius the principal \textbf{right}‐eigenvector of $P$ (the one for eigenvalue $1$) may be chosen with unit Euclidean norm $\|v\|_2=1$,
    \[ v = \frac1{\sqrt{2n}}(1,1,\dots,1)^T . \]
Rewriting with respect to $V_A$ and $V_B$ in \Cref{eqn:vavb_eigvecs}, we get
    \[ v = \onesqrttwo \begin{pmatrix} V_A \\ V_B \end{pmatrix}, \]
coinciding with the balanced quantum‐like state $\kp=\onesqrttwo(\kz+\ko)$ attained when $k_A=k_B$ and $l \neq 0$.

\subsection{Undirected Detuning}
When $G_R$ is not regular but its two halves are $k_A$- and $k_B$-regular and connected by an $l$-regular bipartite matrix $C$, node degrees become
    \begin{equation*}
        \deg(i)=
        \begin{cases}
            k_A+l, & i\in A,\\
            k_B+l, & i\in B.
        \end{cases}
    \end{equation*}
Thus, the stationary distribution of the walk (defined as the top eigenvector of the transition matrix $P \coloneqq D\inv R$, for diagonal degree matrix $D$) is
    \begin{equation}\label{eq:pi_def}
        \pi_i =\frac{\deg(i)}{\sum_j\deg(j)}
        =\begin{cases}
            \dfrac{k_A+l}{n(k_A+k_B+2l)}, & i\in A,\\
            \dfrac{k_B+l}{n(k_A+k_B+2l)}, & i\in B.
        \end{cases}
    \end{equation}
Introducing $X \coloneqq k_A/l$, $Y \coloneqq k_B/l$, and the detuning parameter $\Delta=(k_A-k_B)/(2l)$ (undirected version, from \Cref{eqn:delta_def}), we obtain
    \begin{equation}\label{eq:pi_param}
        \pi_i =
        \begin{cases}
            \dfrac{X+1}{2n\,(X+1-\Delta)}, & i\in A,\\
            \dfrac{Y+1}{2n\,(Y+1+\Delta)}, & i\in B.
        \end{cases}
    \end{equation}
By contrast, the quantum amplitudes $(a,b)$ depend \emph{only} on $\Delta$ (see \Cref{thm:arb_state_construction,eqn:delta_tuning_eqn}), leaving the global scale $X$ (and hence $Y$) free to tune other metrics--e.g., the spectral gap--without altering $\kpsi$.
\footnote{For example, given $\Delta = 0.5$, there are a countably infinite number of ratios $X = \frac{k_A}{l}$ that satisfy $0.5 = \frac{k_A-k_B}{2l} \implies l = k_A - k_B$, such as $n=40,\,X=10$ (or $20$) $\implies \pi_i = 0.0131$ (or $0.0128$).}

\subsection{Directed coupling}
If the bipartite connector is directed with degrees $(l_A,l_B)$ and we set $k_A = k_B$, as in \Cref{sec:asym}, the same logic gives
    \begin{equation}
        \pi_i =
        \begin{cases}
            \dfrac{k+l_A}{n(2k+l_A+l_B)}, & i\in A,\\[3pt]
            \dfrac{k+l_B}{n(2k+l_A+l_B)}, & i\in B,
        \end{cases}
    \end{equation}
where again fixing the quantum amplitudes $(a,b)$ pins down only the ratio $\Delta_C = l_A/l_B$ (see \Cref{lem:asym_coup_eigvecs,eqn:delta_c_tuning_eqn}), leaving the global scales $X \coloneqq l_A/k$ and $Y \coloneqq l_B/k$ free.

\subsection{Summary and Outlook}
Matching a target quantum‐like state $\kpsi$ fixes certain \emph{relative} degree ratios (e.g., $\Delta$ or $\Delta_C$) while leaving a continuous manifold of absolute scalings $(X,Y)$ at one’s disposal.  The classical stationary distribution is sensitive to those scalings and therefore encodes extra information beyond $\ket{\psi}$—information that can be leveraged to optimize spectral separation, robustness, or hardware implementability in multi‐bit extensions (see \Cref{sec:extensions:two_qubits}).

\section{Real-Valued Bipartite Connection Matrix \texorpdfstring{$C$}{C}}\label{sec:real_valued_C}
Allowing real edge weights, in short, makes construction of an arbitrary QL-bit state trivial.  Furthermore, guided by the underlying motivation of in- and out-of-phase oscillators in complex synchronized networks, we find that real edge weights lack physical motivation. Generally, two oscillators are in-sync or out-of-sync, with anything in between being ill-defined and unstable.

We demonstrate the triviality by allowing a minimal relaxation, $[C]_{ij} \in [0, 1]$ for symmetric coupling (for simplicity we assume positive edge weights, but as discussed previously, this is without loss of generality).  This, in turn, allows the ``regularity'' of $C$ to be continuous, $l \in [0, n]$, which allows arbitrary amplification or suppression of the symmetric regularity gap $k_A-k_B$, allowing $\Delta$ in the discrete range $\Delta \in \{0, 1/n, 2/n, \dots, 1\}$ and the continuous range $\Delta \in [1, \infty)$.  Given $\Delta\inv$ (discussed in \Cref{sec:asym:switching_sym}) and arbitrary choice of assignment of $a, b$, this trivially continuously covers all valid states with $|a|^2+|b|^2 = 1$.

\section{Framework Extensions} \label{sec:extensions}
Beyond real-valued extensions, it is also interesting to study graphs with complex weights, as that is presently the primary distinction between quantum-like and quantum bits.  We propose allowing edge weights that are the Gaussian-integer (i.e. complex integer) roots of unity, $e \in \{\pm 1, \pm i\}$.  This would allow ``regularities'' to be complex, affecting what values we are able to represent in $\Delta$, the tuning parameter related to the amplitudes of $\kpsi$.  Similar to \Cref{sec:symm:constraints}, this would allow amplitude ratios $\Delta \in \Q(i)$, for the Gaussian rationals $\Q(i) = \{a + bi : a, b \in \Q\}$. For the same reason discussed in \Cref{sec:real_valued_C}, allowing continuous complex edge weights yields trivial exact representations of quantum states.  Furthermore, even taking the next natural step above the (Gaussian integer) fourth roots of unity $\mu_4$ to the eighth roots of unity $\mu_8 = \{\pm 1, \pm i, \onesqrttwo(\pm1 \pm i)\}$ introduces an unrepresentable number in current classical models of computation, $\sqrt{2}$. We therefore propose the Gaussian integers as the natural complex extension to QL-bit modeling.  With some preliminary work, the authors propose the following symmetric tuning model with complex weights added to the connection matrix $C$:

\begin{conjecture}[Gaussian Integer Edge Weights Yield Complex Approximations of Quantum States]
    Given a new tunable parameter $\Delta_G$ incorporating Gaussian integer edge weights $\left(\mu_4 = \{\pm 1, \pm i\}\right)$ into the connection matrix $C$, defined as
        \[ \Delta_G \coloneqq \frac{k_A-k_B}{2l\eta} \text{ for } \eta \in \mu_4, \]
    \Cref{thm:arb_state_construction} holds with the appropriate replacements and the notable substitution of \Cref{eqn:delta_tuning_eqn} with
    \begin{equation}\label{eqn:delta_g_tuning_eqn}
        \Delta_G = \frac{1}{a^2-b^2} = \frac{1}{2\omega_1\omega_2}
    \end{equation}
    and \Cref{eqn:arb_R_eigval} with 
    \begin{equation}\label{eqn:arb_R_eigval_complex_C}
        \lambda_R = k_A - \frac{\omega_2}{\omega_1}l\eta = k_B - \frac{\omega_1}{\omega_2}l\eta\inv .
    \end{equation}
\end{conjecture}

Additionally, it is interesting to study what single-QL-bit unitary transformations would look like in this framework. We have demonstrated the direct construction of arbitrary QL-bits, but have yet to demonstrate how to reach a target state from an initial state, say $\kpsi = \kp$. Such transformations were implied abstractly in \cite{amati2025qlbits} via graph tensor products. It would be interesting to realize them while maintaining the same number of nodes and the same general graph properties; explicitly, this would take the construction $\kpsi = \kp$ in \Cref{lem:basic_comp_eigvecs} to $\kpsi = a\kp  + b\km$ in \Cref{thm:arb_state_construction}.

\subsection{Relation to gate-model and networked quantum computing.}
On the invariant subspace $\mathcal{S}=\operatorname{span}\{\kz,\ko\}$, the restriction of $R$ is represented in the ordered basis $(\kz,\ko)$ by
    \[ R_{\mathrm{eff}} \coloneqq \bigl[R|_{\mathcal{S}}\bigr]_{(\kz,\ko)}
                = \begin{pmatrix} k_A & -l \\[1mm] -l  & k_B \end{pmatrix}
                = \bar{k}I+l(\sigma_z\Delta-\sigma_x), \qquad \bar{k}\coloneqq\frac{k_A+k_B}{2}, \]
which is equivalently the quotient matrix associated with the equitable partition $V(G_R)=V(G_A)\sqcup V(G_B)$, with entries given by the corresponding signed block-row sums \cite{godsil2013algebraic}. Here $\sigma_x$ and $\sigma_z$ denote the Pauli matrices. Since scalar shifts and nonzero rescalings do not change eigenvectors, $R_{\mathrm{eff}}$ has, at the level of its eigenstates, the same form as a real single-qubit operator in the $x-z$ plane.

Under the identification of $\kz$ and $\ko$ with the computational basis states, a target state $\omega_1\kz+\omega_2\ko$ can be prepared on a gate-model processor, up to a global phase, by the standard $y$-axis rotation
    \[ \omega_1\kz+\omega_2\ko = R_y(\theta)\kz, \qquad \theta=2\operatorname{atan2}(\omega_2,\omega_1), \]
which is the single-qubit instance of standard rotation-based quantum-state preparation \cite{mottonen2005transformation}. For the symmetric detuning construction with nonzero amplitudes, \Cref{eqn:delta_tuning_eqn} then gives $\Delta=-\cot\theta$. In this sense, the present graph construction supplies a classical parameterization of state-preparation targets and effective two-level operators, rather than a replacement for physical quantum gates. This viewpoint may be useful in studies of variational training and adaptive control \cite{gyongyosi2019training,gyongyosi2022adaptive, mcclean2016theory}, as well as modular or distributed gate-model architectures \cite{gyongyosi2021scalable,monroe2014large}.

At the network level, the regular subgraphs and connector blocks give a coarse-grained description of computational modules and inter-module couplings, which may be useful as an abstraction for topology-aware resource studies and the orchestration of networked quantum services \cite{gyongyosi2025networked,gyongyosi2022advances,wehner2018quantum}. The present adjacency-eigenvector model does not, however, include physical noise, entanglement-generation rates, repeater operations, or quantum-channel capacities; those ingredients must be incorporated before quantitative quantum-network performance claims can be made \cite{gyongyosi2018survey}. The asymmetric construction of \Cref{sec:asym} is generally non-Hermitian and should likewise not be identified with a closed-system gate-model Hamiltonian without an additional Hermitian embedding or dilation.

These gate-model and networking interpretations motivate the most natural extension of this work: the two-QL-bit setting.

\subsection{Two QL-Bits}\label{sec:extensions:two_qubits}
Thus far, we have shown explicitly how to construct both a perfect superposition (affirming previous observations) and an arbitrary single-qubit state $\kpsi$. We provide the following intuition for how to extend these efforts to two-qubit states. Previously, \cite{amati2025qlbits} proposed this using graph Cartesian products, and indeed showed how this mathematical mapping extends beyond two qubits, but here we propose constructing the two-qubit state by taking two independent QL-bit graphs and connecting them in such a way as to directly create any two-qubit state.  This method has the advantage of being more intuitive but has the disadvantage of lacking a natural three (or higher) QL-bit extension.

\begin{figure}
    \centering
    \includegraphics[width=1.05\textwidth]{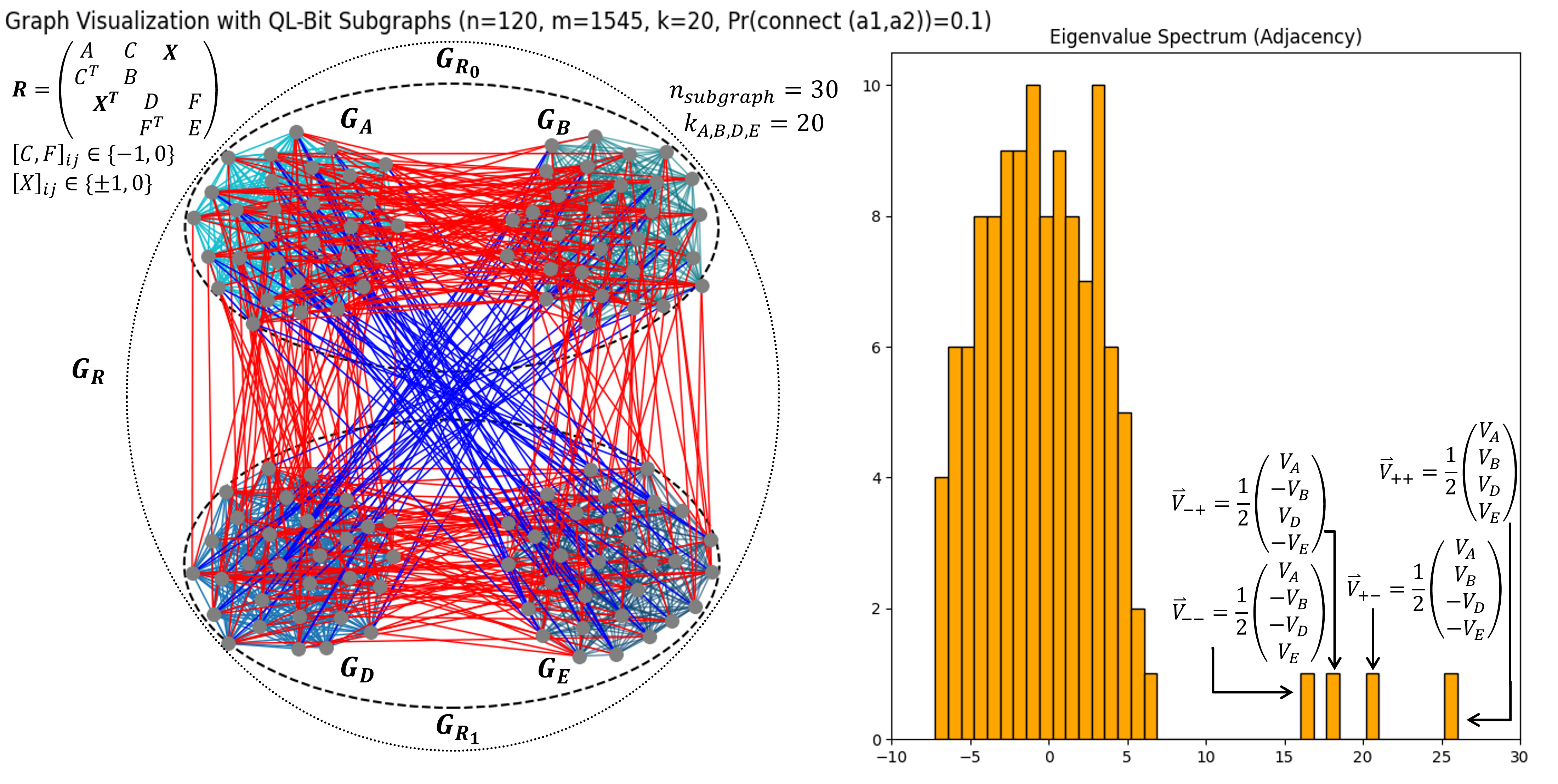}
    \caption{An example of how to connect two QL-bits in the method of \cite{scholes2024quantum}, along with its adjacency spectrum.  See \Cref{fig:graphandspectrum_adj_msubgraphNone} for the single-QL-bit construction. Each subgraph $G_A, G_B, G_D, G_E$ has 30 nodes and is $k=20$ regular internally.  The probability of connection between $G_A \leftrightarrow G_B$ and $G_D \leftrightarrow G_E$ is 0.1 (regularity inside one QL-bit of 20+3), and between the opposite subgraphs it is 0.05 (average regularity between QL-bits of 1.5+1.5). Red lines indicate negative edges between $A\leftrightarrow B, A\leftrightarrow D, B\leftrightarrow E, D\leftrightarrow E$, and blue edges indicate positive between the others. See text for more details.}
    \label{fig:two_qubits_graphandspectrum_adj_msubgraphNone}
\end{figure}

As in \cite{scholes2024quantum}, let $G_R$ be a graph represented by its adjacency matrix $R$, as illustrated in \Cref{fig:two_qubits_graphandspectrum_adj_msubgraphNone}.  In this construction, two QL-bits $R_0, R_1$ are built from four subgraphs $A, B \in R_0,  D, E \in R_1$, as expected, with internal bipartite connection matrices $C \in R_0, F \in R_1$.  To interact the two QL-bits, we use 4 cross-QL-bit bipartite connection matrices $X_{AD}, X_{AE}, X_{BD}, X_{BE} \in X$ to connect the individual subgraphs of $R_0$ and $R_1$.

This defines the (undirected) block adjacency matrix
    \begin{align}
        R = \begin{pmatrix} R_0 & X \\[1mm] X^T & R_1 \end{pmatrix}, \quad X \coloneqq \begin{pmatrix} X_{AD} & X_{AE} \\[1mm] X_{BD} & X_{BE} \end{pmatrix},
    \end{align}
or, symbolically (and directed),
    \begin{align}
        R = \left(
        \begin{array}{cc|cc}
            A       & A \to B & A \to D & A \to E \\[1mm]
            B \to A & B       & B \to D & B \to E \\[1mm] \hline
            D \to A & D \to B & D       & D \to E \\[1mm]
            E \to A & E \to B & E \to D & E
        \end{array} \right),
    \end{align}
where $A\rightarrow B$ represents edges from $A$ to $B$.

When $R_0, R_1, X$ are chosen to be regular, we propose that one may use the corresponding quantum-like basis vectors (as defined in \Cref{eqn:kp_km}) to create the perfect superposition state
    \begin{equation}
        \psi_{++} \equiv \onesqrttwo(\kp + \kp)
        = \frac12 \left( \begin{pmatrix} V_A \\[1mm] V_B \\[1mm]
                                         \mathbf{0}_n \\[1mm]
                                         \mathbf{0}_n \end{pmatrix} +
                         \begin{pmatrix} \mathbf{0}_n \\[1mm]
                                         \mathbf{0}_n \\[1mm]
                                         V_D \\[1mm]
                                         V_E \end{pmatrix}
                  \right)
        = \frac12 \begin{pmatrix} V_A \\[1mm] V_B \\[1mm] V_D \\[1mm] V_E \end{pmatrix}
    \end{equation}
with appropriate eigenvalue $\lambda_{++} = (k+l)+(j_1+j_2)$ (for $k$-regular subgraphs $R_0$ and $R_1$, $l$-regular internal qubit connection graphs $C$ and $F$, and $j$-regular qubit-to-qubit connection graph $X$). This formula is very opaque, but $j_1$ represents all of the edges going from, for example, $A \to D$ and $j_2$ represents $A \to E$; then the sum $j_1+j_2$ is the (average block) sum of row 1 of $X$.  Similarly, such a construction would create eigenvectors
\begin{align*}
    \psi_{+-} \equiv \onesqrttwo (\kp - \kp) = \frac12 &\begin{pmatrix} V_A \\[1mm] V_B \\[1mm] -V_D \\[1mm] -V_E \end{pmatrix},\quad
    \psi_{-+} \equiv \onesqrttwo (\km + \km) = \frac12 \begin{pmatrix} V_A \\[1mm] -V_B \\[1mm] V_D \\[1mm] -V_E \end{pmatrix},\text{ and}\\
    \psi_{--} \equiv& \onesqrttwo (\km - \km) = \frac12 \begin{pmatrix} V_A \\[1mm] -V_B \\[1mm] -V_D \\[1mm] V_E \end{pmatrix} ,
\end{align*}
with corresponding eigenvalues $\lambda_{+-} = (k+l)-(j_1+j_2),  \lambda_{-+} = (k-l)+(j_1-j_2),  \lambda_{--} = (k-l)-(j_1-j_2)$, as illustrated in \Cref{fig:two_qubits_graphandspectrum_adj_msubgraphNone}.

We pause to note here that we are (abusively) mapping the algebraic tensor product $\kpp$ to the summation of zero-extended basis vectors, as was done in \Cref{eqn:kp_km}. Whether this is the correct extension is open and being studied by the authors, but we know that as the graph expands to $N$ QL-bits, the block adjacency matrix must expand as $2N$.  Therefore the eigenvectors, which map to the quantum state, also grow as $2N$.

The goal with two QL-bits is to construct an emergent eigenvector of $R$ corresponding to the arbitrary state
    \[ \kpsi = a\psi_{++} + b\psi_{+-} + c\psi_{-+} + d\psi_{--} \quad \text{with} \quad |a|^2+|b|^2+|c|^2+|d|^2 = 1 . \]

\section{Conclusion}
We have built upon the framework developed in \cite{scholes2023largecoherentstates,scholes2024quantum,scholes2025productstates} that defined and showed how to manipulate QL-bits. In particular, we provide mathematically rigorous methods to consistently create arbitrary single QL-states 
    \[ \psi = a\,\psi_{+}+b\,\psi_{-},\quad |a|^2+|b|^2=1 \]
without need for ensembles of states. In \Cref{sec:sym}, we showed how to create these by manipulating only subgraph regularities, $\Delta=\frac{k_A-k_B}{2l}$.  In \Cref{sec:asym}, we showed how to do the same by allowing the connecting edges to be directed and manipulating only those regularities, $\Delta_C = \frac{l_A}{l_B}$.  In both cases, to prevent unbounded behavior, the introduction of $\Delta\inv$ and $\Delta_C\inv$ was necessary along with switching conditions for using each.  Finally, in \Cref{sec:random_walks} we provide an explicit connection between the QL-bit states and random walks on graphs.

Future research opportunities in this area remain broad.
In \Cref{app:spect_gap}, we discuss graph parameters versus spectral gap, and assume spectral gap is known to be constant for distinguishability purposes.
While this is sufficient for single QL-bits, many-body systems or systems required to evolve will require more robust spectral guarantees to ensure top eigenvalues remain distinct in the spectrum.

This work is the first to show creation of arbitrary single QL states that scales with the desired precision of the ratio of the amplitudes (as a fraction does to real numbers).  Other works have only shown construction of equal superposition in expectation \cite{scholes2024quantum} or graph transformations that require taking the Cartesian product of two graphs, an operation which scales exponentially with the number of desired operations and itself requires real edge weights (which this work does not).  The authors are currently working on if these techniques can be extended to gates or even single-QL-bit unitaries in a similarly rigorous fashion, as discussed in \Cref{sec:extensions}.

Given the tight similarities between these QL states and steady states of random walks on a graph, many further directions could be explored from a computer science perspective, in much of the same way as random walks have been explored. While the current construction of different eigenstates does allow us to correlate with the steady state distribution due to the dependence on the diagonal matrix $D$, an alternate construction that directly manipulates $D^{-1}R$ in order to create state may allows us to better relate the eigenstates to steady states.

Finally, with such a physically fundamental natural concept as spontaneous non-linear synchronization of complex synchronized networks in nature, it could be interesting to see if these networks are performing quantum-like computation at a macro scale.  Researchers have long wondered about how quantum processes could drive neurological biological processes \cite{beck1998quantum,adams2020quantum,jedlicka2017revisiting,tegmark2000importance}, this framework could give insights from a more classical perspective.

\section*{Acknowledgements}
We acknowledge Prof. Gregory D Scholes for insightful discussions.  S.K. acknowledges funding from the Office of Science through the Quantum Science Center (QSC), a National Quantum Information Science Research Center.\\
\vskip6pt

 \newpage
 {\bf {\Large Appendices:}}
\appendix
\section{``Large Enough'' Spectral Gap}\label{app:spect_gap}
Continuing from the discussion in \Cref{sec:intro}, we analyze the spectral gap (\Cref{eqn:spec_gap_ER_graphs})
    \[ \lambda_1 - \lambda_2 \ge np - 2\sqrt{np(1-p)} .\]
The aim of this section is to understand practically when a spectral gap becomes ``large enough'' to be distinguishable as a parameter of $n$ and $p$.

\begin{figure}
    \centering
    \includegraphics[width=0.5\linewidth]{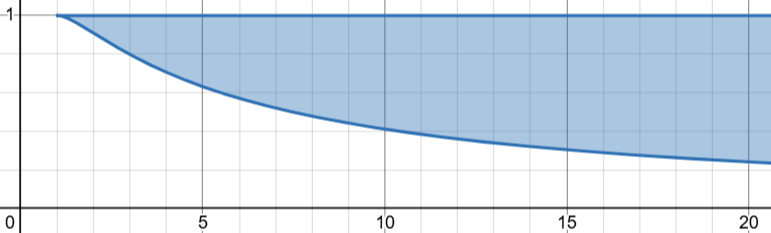}
    \caption{Spectral gap feasible region as a function of graph size $n$ (x-axis) and edge existence probability $p$ (y-axis) for spectral gaps $\lambda_1 - \lambda_2 \ge 1$.  Shows minimum graph size as a function of probability of connection; $k$-regularity is defined as $k=np$, which is $\lambda_1$ w.h.p.
    Reproducible in \Cref{app:fancy_plot_eqns}.}
    \label{fig:spectral_gap_feasible_n_vs_p}
\end{figure}

Arbitrarily setting a minimum spectral gap
    \[ |\lambda_1-\lambda_2| \ge 1 ,\]
we get the set of constraints
    \begin{equation}\label{eqn:spectral_gap_bounds}
        1 \le np - 2\sqrt{np(1-p)}\,, \quad 0 \le p \le 1 \,.
    \end{equation} 
Visually, we can see these constraints in \Cref{fig:spectral_gap_feasible_n_vs_p}.  Solving \Cref{eqn:spectral_gap_bounds} for $p$ yields the feasible solution:
    \begin{equation*}
        2\sqrt{\frac{2n-1}{n(n+4)^2}} + \frac{3}{n+4} \le p \le 1 \quad \text{for } n \ge 1 \,.
    \end{equation*}
Where $n=1 \implies p=1$. Equivalently solving for $n$:
    \begin{equation}\label{eqn:min_n_given_pr_conn}
        n \ge \frac{6+2\sqrt{p^2-3p+2}}{p}-2 \text{ for } 0<p \le 1 \,.
    \end{equation}
If $p=0$, there is no graph (the entire matrix is zeros) and therefore no spectral gap. \Cref{eqn:min_n_given_pr_conn} tells us the minimum $n$ given a probability of connection $p$, as visualized in \Cref{fig:spectral_gap_feasible_n_vs_p}.

Roughly, these equations give the intuition that the regularity $k$ of a graph should not be \textit{too small} compared to the overall graph size $n$.  A big enough $k$ ensures that each QL-bit's emergent eigenvalues are easily distinguishable (and therefore robust for quantum-like computation).

Precisely, for an arbitrary gap $a$ yielding
    \begin{equation}\label{eqn:p_range_spectral_gap_generic}
        2\sqrt{\frac{-a^2+an+n}{n(n+4)^2}} + \frac{a+2}{n+4} \le p \le 1 \quad \text{for } n \ge a > 0 \,,
    \end{equation}
we have the following trivial lemma.

\begin{lemma}[Minimum Regularity]\label{lem:min_regularity}
    Given a random regular graph of size $n \ge 1$ and edge existence probability $p$ (\ER graph), the minimum average regularity $k = np$ (defined by the minimum $p$) to achieve a minimum spectral gap of $|\lambda_1 - \lambda_2| \ge a$, for constant $a$, as $n \to \infty$ is constant,
    \begin{equation}
        \lim_{n\to\infty} np = 2 + a + 2\sqrt{1+a} \, .
    \end{equation}
    \begin{proof} Follows trivially from limits and \Cref{eqn:p_range_spectral_gap_generic}.
        \begin{align*}
            \lim_{n\to\infty} np
                &\stackrel{(\ref{eqn:p_range_spectral_gap_generic})}{=} \lim_{n\to\infty} n\left(2\sqrt{\frac{-a^2+an+n}{n(n+4)^2}} + \frac{a+2}{n+4}\right)\\
                &= \lim_{n\to\infty} 2n\sqrt{\frac{-a^2+an+n}{n(n+4)^2}} + \lim_{n\to\infty}\frac{an+2n}{n+4}\\
                &= \lim_{n\to\infty} 2n\sqrt{\frac{-a^2/n+a+1}{n^2(1+\frac4n)^2}} + \lim_{n\to\infty}\frac{a+2}{1+\frac4n}\\ %
                &= \left(\lim_{n\to\infty} 2\frac{\sqrt{-a^2/n+a+1}}{1+\frac4n}\right) + a+2\\
                &= 2\sqrt{a+1} + a+2
        \end{align*}
    \end{proof}
\end{lemma}

This tells us that for any graph size and desired (constant) spectral gap, a random regular graph needs merely constant average regularity of order directly proportional to the gap, $O(a+\sqrt{a}) = O(1)$, to have a spectral gap $\ge a$.  For example, if an experimentalist indicates they need a minimum spectral gap of $1$ to use these graphs for quantum computation, this says that they need a regularity of $k \ge 6$ no matter the graph size $n$.  For non-constant spectral gaps $A(n)$, loosely speaking, if $A(n)$ converges, the limit converges as above, if $A(n) \in o(n)$ grows slower than $n$, the limit converges to the limit of $A(n)$ similarly as above, and if $A(n)$ grows similarly or faster than $n$, the limit either diverges or becomes non-real.  In short, the minimum regularity is tightly proportional to the desired spectral gap (as expected).
Current simulations of quantum computation using QL-bits only use a constant spectral gap \cite{amati2025qlbits}.

\section{Reconstructing Plots with Equations}\label{app:fancy_plot_eqns}
For \Cref{fig:basic_visualization,fig:full_visualization}:
\begin{verbatim}
    //Made with Desmos 3D (https://www.desmos.com/3d/rh6ae2lkb1)
    //Settings --> reverse contrast and translucent surfaces.
    x^{2}+y^{2}=1 [extend to 3D]
    z\ =\ \frac{2xy}{y^{2}-x^{2}}\left\{\ x^{2}+y^{2}=1\right\}
    x\ =\ \sqrt{\frac{1}{2}\left(1+\frac{1}{\sqrt{z^{2}+1}}\right)}
    x\ =\ \sqrt{\frac{1}{2}\left(1-\frac{1}{\sqrt{z^{2}+1}}\right)}
    x\ =\ -\sqrt{\frac{1}{2}\left(1+\frac{1}{\sqrt{z^{2}+1}}\right)}
    x\ =\ -\sqrt{\frac{1}{2}\left(1-\frac{1}{\sqrt{z^{2}+1}}\right)}
    z = 1
    z = 2
    z = -1
    z = -2
\end{verbatim}
To add visualization of $\Delta\inv$ beyond \Cref{fig:basic_visualization_delta_delta_inv}, as discussed in \Cref{sec:asym:switching_sym}, add the following equations to the above equations:
\begin{verbatim}
    //(https://www.desmos.com/3d/kzaoepp3j2)
    z\ =\ \frac{y^{2}-x^{2}}{2xy}\left\{\ x^{2}+y^{2}=1\right\}
    x\ =\ \sqrt{\frac{1}{2}\left(1+\frac{z}{\sqrt{z^{2}+1}}\right)}
    x\ =\ \sqrt{\frac{1}{2}\left(1-\frac{z}{\sqrt{z^{2}+1}}\right)}
    x\ =\ -\sqrt{\frac{1}{2}\left(1+\frac{z}{\sqrt{z^{2}+1}}\right)}
    x\ =\ -\sqrt{\frac{1}{2}\left(1-\frac{z}{\sqrt{z^{2}+1}}\right)}
\end{verbatim}

\noindent For \Cref{fig:basic_visualization_delta_c,fig:full_visualization_delta_c}
\begin{verbatim}
    //Made with Desmos 3D (https://www.desmos.com/3d/5dcghnbitv)
    //Settings --> reverse contrast and translucent surfaces.
    x^{2}+y^{2}=1 [extend to 3D]
    //delta_C
    z\ =\ \left(\frac{x+y}{x-y}\right)^{2}\left\{\ x^{2}+y^{2}=1\right\}
    //delta_C inverse
    z\ =\ \left(\frac{x-y}{x+y}\right)^{2}\left\{\ x^{2}+y^{2}=1\right\}
    x\ =\ \sqrt{\frac{1}{2}+\frac{\sqrt{z}}{z+1}}
    x\ =\ \sqrt{\frac{1}{2}-\frac{\sqrt{z}}{z+1}}
    x\ =\ -\sqrt{\frac{1}{2}+\frac{\sqrt{z}}{z+1}}
    x\ =\ -\sqrt{\frac{1}{2}-\frac{\sqrt{z}}{z+1}}
    z = 1
    z = 2
    z = -1
    z = -2
\end{verbatim}

\noindent For \Cref{fig:spectral_gap_feasible_n_vs_p}:
\begin{verbatim}
    //Made with Desmos (https://www.desmos.com/calculator/ng8ckmfa0b)
    2\ \sqrt{\frac{2x-1}{x\left(x+4\right)^{2}}}
     + \frac{3}{x+4}\le y\le1\left\{x>1\right\}
    OR
    x\ge2\ \sqrt{\frac{y^{2}-3y+2}{y^{2}}}
          +\frac{3}{y}-2  \left\{0<y\le1\right\}
\end{verbatim}
See also: \url{https://www.wolframalpha.com/input?i=solve+1\%3C\%3D+xy-2\%5Csqrt\%7Bxy\%281-y\%29\%7D+for+0\%3C\%3Dy\%3C\%3D1} and \url{https://www.wolframalpha.com/input?i=solve+1\%3C\%3D+xy-2\%5Csqrt\%7Bxy\%281-y\%29\%7D\%2C+0\%3C\%3Dy\%3C\%3D1+for+x}.

\setcounter{equation}{0}
\renewcommand{\theequation}{C \arabic{equation}}
\section{General Characteristic Equation of \texorpdfstring{$G_R$}{G\_R}}\label{app:general_characteristic_eqn}
This section generalizes the analysis done for \Cref{thm:arb_state_construction} and \Cref{lem:asym_coup_eigvecs}.

\begin{lemma}\label{lem:characteristic_eqn_satisfies_eigvec}
    If we allow $G_R$ to be constructed as before out of two regular subgraphs $G_A$ and $G_B$, with regularities $k_A,\, k_B$, each of order $n$, with entries $[A]_{ij}\,, [B]_{ij} \in\{0, 1\}$, coupled by bipartite graph $G_C$ with $l_A,\,l_B$-regular \emph{directed} connection matrices $C_A, C_B$ (with $[C_A]_{ij},\,[C_B]_{ij}\in\{-1,0\}$), then, to ensure
        \[ \kpsi \coloneqq V_R = a\,\kp+b\,\km = \begin{pmatrix} \omega_1 V_A \nonumber\\[1mm] \omega_2 V_B \end{pmatrix} \]
    (\Cref{eqn:arb_qubit_def}) is an eigenvector of 
        \[ R=\begin{pmatrix} A & C_A \\[1mm] C_B & B \end{pmatrix} ,\]
    $k_A, k_B, l_A, l_B, \omega_1, \omega_2$ must satisfy
        \begin{equation}\label{eqn:general_characteristic_eqn}
            \omega_2^2 \,l_A - \omega_1^2 \,l_B + \omega_1\omega_2 \, (k_B-k_A) = 0 .
        \end{equation}
    by the eigenvector equation. This differs from the lemma and theorem in that all 4 regularities ($k_A, k_B, l_A, l_B$) are allowed to differ.
\end{lemma}

\begin{proposition}[Generalization of Symmetric and Asymmetric Coupling for Arbitrary QL-Bits]
    (a) \Cref{thm:arb_state_construction} and (b) \Cref{lem:asym_coup_eigvecs} are special cases of \Cref{eqn:general_characteristic_eqn}.

    \begin{proof}
        \begin{enumerate}
            \item[(a)] As in \Cref{thm:arb_state_construction}, take $l_A = l_B$, then \Cref{eqn:general_characteristic_eqn} becomes
                \[ (\omega_2^2 - \omega_1^2)l + \omega_1\omega_2 \, (k_B-k_A) = 0 , \]
            yielding
                \[ \frac{k_A-k_B}{l} = \frac{\omega_2^2-\omega_1^2}{\omega_1\omega_2}, \]
            resulting in \Cref{eqn:delta_def,eqn:delta_tuning_eqn}.
    
            \item[(b)] As in \Cref{lem:asym_coup_eigvecs}, take $k_A = k_B$, then \Cref{eqn:general_characteristic_eqn} becomes
                \[ \omega_2^2 \,l_A - \omega_1^2 \,l_B = 0 , \]
            yielding
                \[ \frac{\omega_1^2}{\omega_2^2} = \frac{l_A}{l_B}, \]
            resulting in \Cref{eqn:delta_c_def,eqn:delta_c_tuning_eqn}.
        \end{enumerate}
    \end{proof}
\end{proposition}

This shows that an arbitrary QL-bit can be constructed by two regular subgraphs joined by two regular directed bipartite connection matrices, as long as the regularities $k_A, k_B, l_A, l_B$ are related to the quantum state amplitudes $\omega_1, \omega_2$ by \Cref{eqn:general_characteristic_eqn}.

\newpage
\bibliographystyle{abbrv} %
\bibliography{refs} %

\end{document}